\documentclass[conference]{IEEEtran}

\usepackage{tikz}
\usetikzlibrary{tikzmark}
\usetikzlibrary{patterns}

\usepackage{xspace}
\usepackage{multirow}
\usepackage{balance}
\usepackage{ctable}
\usepackage{listings}
\usepackage{color}
\usepackage{subcaption}
\usepackage{graphicx}
\graphicspath{ {figures/} }
\usepackage{url}
\usepackage[justification=centering]{caption}

\usepackage{algorithm}
\usepackage{algorithmicx}
\usepackage{algpseudocode}

\usepackage{amssymb}
\usepackage{amsmath}
\usepackage{amsfonts}

\usepackage{newfloat}

\usepackage{cleveref}
\usepackage{acronym}
\usepackage{verbatim} 

\usepackage{xcolor}

\usepackage[nounderscore]{syntax}
\usepackage{mdframed}

\usepackage{multicol}
\usepackage{multirow}
\usepackage[flushleft]{threeparttable}

\usepackage{enumitem}

\usepackage{array, makecell, caption}
\usepackage{cellspace}

\usepackage{pifont}
\usepackage{tcolorbox}

\definecolor{brown}{cmyk}{0,0.81,1,0.60}
\definecolor{magenta}{rgb}{0.4,0.7,0}
\definecolor{gray}{rgb}{0.5,0.5,0.5}
\definecolor{red}{rgb}{1,0,0}
\definecolor{green}{rgb}{0.5,0,0.5}
\definecolor{blue}{rgb}{0,0,1}

\newcommand{\ie}{i.e.,\xspace}
\newcommand{\eg}{e.g.,\xspace}

\definecolor{mpcolor}{rgb}{0.9,0.45,0.1}

\definecolor{smcolor}{rgb}{0.1,0.1,0.7}

\definecolor{todocolor}{rgb}{1,0.3,0.3}

\renewcommand{\paragraph}[1]{\vskip 0.05in \noindent\textbf{#1.}}

\newcommand{\RQ}[2]{\vspace{3pt}\noindent\textbf{#1:} #2}

\newfloat{program}{h}{eqn}
\newlength{\MaxSizeOfLineNumbers}%
\definecolor{keywordcolor}{rgb}{0.8,0.1,0.5}
\definecolor{lightlightgray}{gray}{.96}
\definecolor{lightgray}{gray}{.85}
\definecolor{medlightgray}{gray}{0.7}
\definecolor{medgray}{gray}{0.4}
\definecolor{darkgray}{gray}{0.35}
\definecolor{nearblack}{gray}{0.15}

\newcommand{\code}[1]{{\small\texttt{#1}}}

\crefname{program}{Program}{Programs}

\algnewcommand\algorithmicswitch{\textbf{switch}}
\algnewcommand\algorithmiccase{\textbf{case}}
\algdef{SE}[SWITCH]{Switch}{EndSwitch}[1]{\algorithmicswitch\ #1\ \algorithmicdo}{\algorithmicend\ \algorithmicswitch}%
\algdef{SE}[CASE]{Case}{EndCase}[1]{\algorithmiccase\ #1}{\algorithmicend\ \algorithmiccase}%
\algtext*{EndSwitch}%
\algtext*{EndCase}%

\definecolor{javared}{rgb}{0.6,0,0} 
\definecolor{javagreen}{rgb}{0.25,0.5,0.35} 
\definecolor{javapurple}{rgb}{0.5,0,0.35} 
\definecolor{javadocblue}{rgb}{0.25,0.35,0.75} 
\definecolor{pastelyellow}{rgb}{0.99, 0.99, 0.59}
\definecolor{peach-orange}{rgb}{1.0, 0.8, 0.6}
\definecolor{peach}{rgb}{1.0, 0.9, 0.71}
\definecolor{lightgoldenrodyellow}{rgb}{0.98, 0.98, 0.82}
\definecolor{lightgreen}{HTML}{CCFFCC}
\definecolor{magicmint}{rgb}{0.67, 0.94, 0.82}
\definecolor{lightmauve}{rgb}{0.86, 0.82, 1.0}
\definecolor{palepink}{rgb}{0.98, 0.85, 0.87}
\definecolor{lightapricot}{rgb}{0.99, 0.84, 0.69}
\definecolor{blizzardblue}{rgb}{0.67, 0.9, 0.93}
\definecolor{mauve}{HTML}{E0D7FF}
\definecolor{orange}{HTML}{FFE5B4}
\definecolor{blue}{HTML}{BFEFFF}
\definecolor{githubGreen}{HTML}{D7FEDA}
\definecolor{githubRed}{HTML}{F9DDE0}
\definecolor{textRed}{HTML}{C11B17}
\definecolor{textBlue}{HTML}{0F52BA}
\definecolor{textOrange}{HTML}{228B22}
\definecolor{textBackgroundGray}{HTML}{E8E8E8}
 
\newcommand{\highlightgray}[1]{%
  \begingroup
  \setlength{\fboxsep}{0.5pt}%
  \colorbox{black!15}{#1}%
  \endgroup
}

\renewcommand\fbox{\fcolorbox{red}{white}}

\newcommand{\cmark}{\ding{52}}%
\newcommand{\xmark}{\ding{56}}%

\makeatletter
\let\orig@lstnumber=\thelstnumber
\newcommand\lstsetnumber[1]{\gdef\thelstnumber{#1}}
\newcommand\lstresetnumber{\global\let\thelstnumber=\orig@lstnumber}
\makeatother

\renewcommand\fbox{\fcolorbox{black}{white}}

\acrodef{apr}[APR]{Automatic Program Repair}
\acrodef{ast}[AST]{Abstract Syntax Tree}
\acrodef{apg}[APG]{Abstract Program Graph}
\acrodef{localization}[FSL]{Failure Scenario Localization}
\acrodef{so}[SO]{\mbox{Stack Overflow}}
\acrodef{re}[RE]{Runtime Exception}
\acrodef{pattern}[REP]{Runtime Exception Pattern}
\acrodef{dsl}[DSL]{Domain Specific Language}

\newcommand{\toolname}{\mbox{\sc Maestro}\xspace}

\newcommand{\maestro}{\mbox{\sc Maestro}\xspace}

\newcommand{\NumBenchmarkSubjects}{78\xspace}
\newcommand{\NumExceptionTypes}{19\xspace}
\newcommand{\NumREPs}{158\xspace} 

\newcommand{\OurUsefulArtifacts}{81\%\xspace}
\newcommand{\OurCorrectPatchesTopOne}{27\%\xspace}
\newcommand{\OurCorrectPatchesTopThree}{40\%\xspace}
\newcommand{\NoREPUsefulArtifacts}{65\%\xspace}
\newcommand{\NoREPCorrectPatchesTopOne}{19\%\xspace}
\newcommand{\NoREPCorrectPatchesTopThree}{22\%\xspace}
\newcommand{\AvgRuntimeInSec}{1\xspace}
\newcommand{\NoREPPatchLoss}{45\%\xspace} 

\AtBeginDocument{%
  \providecommand\BibTeX{{%
    \normalfont B\kern-0.5em{\scshape i\kern-0.25em b}\kern-0.8em\TeX}}}

\begin{document}

\title{Providing Real-time Assistance for Repairing Runtime Exceptions using Stack Overflow Posts}

\author{
    \IEEEauthorblockN{Sonal Mahajan}
    \IEEEauthorblockA{\textit{Fujitsu Research of America, Inc.}\\
    smahajan@fujitsu.com}
    \and
    \IEEEauthorblockN{Mukul R. Prasad}
    \IEEEauthorblockA{\textit{Fujitsu Research of America, Inc.}\\
    mukul@fujitsu.com}
}

\maketitle
\thispagestyle{plain}
\pagestyle{plain}

\begin{abstract}

\acp{re} are an important class of bugs that occur frequently during code development. Traditional \ac{apr} tools are of limited use in this ``in-development" use case, since they require a test-suite to be available as a patching oracle. Thus, developers typically tend to manually resolve their in-development \acp{re}, often by referring to technical forums, such as \ac{so}. To automate this manual process we extend our previous work, \toolname, to provide real-time assistance to developers for repairing Java \acp{re} by recommending a relevant patch-suggesting \ac{so} post and synthesizing a repair patch from this post to fix the \ac{re} in the developer's code. \toolname exploits a library of \acp{pattern}  semi-automatically  mined  from \ac{so} posts, through a relatively inexpensive, one-time, incremental process. An \ac{pattern} is an abstracted sequence of statements that triggers a given \ac{re}. \acp{pattern} are used to index \ac{so} posts, retrieve a post most relevant to the \ac{re} instance exhibited by a developer's code and then mediate the process of extracting a concrete repair from the \ac{so} post, abstracting out post-specific details, and concretizing the repair to the developer's buggy code. We evaluate \toolname on a published \ac{re} benchmark comprised of \NumBenchmarkSubjects instances. 
\toolname is able to generate a correct repair patch at the top position in \OurCorrectPatchesTopOne of the cases, within the top-3 in \OurCorrectPatchesTopThree of the cases and overall return a useful artifact in \OurUsefulArtifacts of the cases. Further, the use of \acp{pattern} proves instrumental to all aspects of \toolname's performance, from ranking and searching of \ac{so} posts to synthesizing patches from a given post. In particular, \NoREPPatchLoss of correct patches generated by \toolname could not be produced by a baseline technique not using \acp{pattern}, \emph{even} when provided with \toolname's \ac{so}-post ranking. \toolname is also fast, needing around \AvgRuntimeInSec second, on average, to generate its output. Overall, these results indicate that \toolname can provide effective real-time assistance to developers in repairing \acp{re}.

\end{abstract}


\section{Introduction}
\label{sec:introduction}

Code search and re-use has long been recognized as a very natural part of software development~\cite{Sadowski:2015,AROMA:OOPSLA2019}. 
Technical discussion forums such as \acf{so}, a rich resource of succinct code artifacts embedded in explanatory text, provide an attractive option to developers for such code search and reuse. 
In fact, software developers frequently visit \ac{so} to resolve issues arising during software development, in particular those related to software debugging and patching~\cite{CrowdDebuggingFSE2015, soUseIEEESoft2017, BeyerESE2020}.

Motivated by the above observation, in this work we propose a technique to provide \emph{real-time} automated support for a developer to resolve a bug in her code by finding and adapting a suitable patch suggested in an \ac{so} post. In particular, we focus on resolving Java \acfp{re}, a common and important class of errors that have attracted significant research in the area of automated debugging and patching~\cite{elixir2017, FuzzyCatch:FSE2020, Sinha:ISSTA2009, NPEFix:arXiv2015, VFix:ICSE2019, GetAFix:OOPSLA:2019, Genesis:FSE2017, Droix:ICSE2018, Long:PLDI2014, Ares:ASE2016, Toradocu:ISSTA2016}, and are also well represented in \ac{so} discussions~\cite{MAESTRO:FSE2020}.  Broadly, our technique shares the patch-generation goal of \acf{apr} techniques~\cite{weimer2009automatically, kim2013automatic, Angelix:ICSE2016, cocount:ISSTA2020, aprCACM2019, aprSurvey2019}. 
However, traditional \ac{apr} techniques, which operate on a \emph{patch synthesis} paradigm and typically rely on a test suite as a patching oracle, involve long running times of tens of minutes or even hours. Hence they are not a good fit for our use case. Developers seeking to resolve an \ac{re}, encountered during development, on \ac{so}, are typically looking for real-time assistance, and may not have a test suite available. 


\paragraph{SO-driven repair \& Challenges} We posit that one solution lies in mimicking human developers' approach -- first \emph{finding} an \ac{so} post discussing a bug similar to theirs and \emph{adapting} the proposed patching solution to their code (in contrast to the \emph{patch synthesis} paradigm of current APR techniques). Such a \emph{find-and-adapt} repair approach would necessarily need to solve three principal challenges 
highlighted in a recent study by Wu et al.~\cite{Wu2019}, namely (C1) the mixed quality of \ac{so} code artifacts, (C2) the difficulty of comprehending code snippets, and (C3) the effort required to modify a chosen snippet to work in their code context. QACrashFix~\cite{QACrashFix:ASE2015} proposes one such approach, for generating patches for Android-related crashes from \ac{so} posts. However, QACrashFix relies on descriptive platform-specific information from Android crashes. As we show in Section~\ref{sec:evaluation}, QACrashFix's approach does not work well for general \acp{re}.

\paragraph{Insight} The guiding insight of our proposed approach is that \emph{bug scenarios} (exception-triggering scenarios here) showcased in \ac{so} post questions, also the root causes for the developers' (exception) bugs, can form the basis for all key steps of a find-and-adapt repair approach -- indexing \ac{so} posts and retrieving a post best matching the bug in a given developer's code (C1), extracting a generalized patch from question and answers in the post (C2), and adapting this patch to the developer's code context (C3). Note that by contrast, current APR techniques are organized around the mining and exploitation of \emph{repair patterns} rather than bug patterns. Our previous work \cite{MAESTRO:FSE2020}, automatically extracts \emph{approximate} bug patterns, termed Exception Scenario Patterns (ESPs), by applying a pre-determined set of abstractions to program statements appearing in both question and answer code snippets of a post. These ESPs are then used as a basis for matching a given RE-generating developer code to a relevant \ac{so} post, to be then \emph{manually} exploited by the developer. However, while such approximate bug patterns are adequate for searching relevant posts, as shown in Section~\ref{sec:evaluation}, they are not \emph{accurate} enough to support automatic patch generation, which is the focus of this work.
 

\paragraph{Approach}
We extend our prior work, \toolname (\textbf{M}ine and \textbf{A}nalyz\textbf{E} \textbf{ST}ackoverflow to fix \textbf{R}untime excepti\textbf{O}ns)~\cite{MAESTRO:FSE2020}, to provide real-time patch-generation support to fix (Java) \acp{re} in developers' code. Our patching technique is built on top of a library of specifications called \acfp{pattern}, semi-automatically mined from \acf{so}, through a relatively inexpensive, one-time, incremental process. Each \ac{pattern} represents an abstract exception-triggering pattern for a specific \ac{re}, discussed in one or more \ac{so} posts. 

Given a developer's code, crashing with an \ac{re}, \toolname first finds a \ac{pattern} from the library best describing the developer's error and using it finds the best \ac{so} post mirroring that error pattern. Then \toolname uses the identified \ac{pattern} to mediate the creation of a complete patch for the developer's buggy code by appropriately re-purposing an answer code snippet in the identified best \ac{so} post. To amplify the efficacy and scope of its patch generation \toolname also uses the \ac{pattern} library to rank the corpus of \ac{so} posts and a novel rule-based re-writing technique to make unparsable \ac{so} snippets parsable. We evaluate \toolname on an existing benchmark of \NumBenchmarkSubjects \ac{re} instances spanning \NumExceptionTypes prominent \acp{re}.
Our evaluation shows that \toolname is able to generate a correct repair patch at the top position in \OurCorrectPatchesTopOne of the cases, within the top-3 in \OurCorrectPatchesTopThree of the cases, and overall return a useful artifact (a correct patch, an almost correct patch, or simply a relevant post) in \OurUsefulArtifacts of the cases. Further, \toolname only required \AvgRuntimeInSec second, on average, per subject, validating its suitability for providing real-time assistance to developers in repairing \acp{re}. The evaluation also shows that the use of \acp{pattern} is key to all aspects of \toolname's performance from ranking and searching of \ac{so} posts to synthesizing patches from a given post. In particular, \NoREPPatchLoss of correct patches generated by \toolname could not be produced by a baseline technique not using \acp{pattern}, \emph{even} when provided with \toolname's \ac{so}-post ranking. 

This paper makes the following contributions:
\begin{itemize}
    \item \textbf{\ac{pattern} Library.} A library of \NumREPs \acp{pattern} spanning \NumExceptionTypes prominent \ac{re} types, systematically mined from \acf{so}, that can be used to characterize the exception-triggering patterns in \ac{so} posts, and hence to index, search and analyze those posts.
    \item \textbf{Patching Technique.} An automated technique and tool \toolname, that employs this \acp{pattern} library to find an \ac{so} post mirroring the error pattern of the \ac{re} in a developer's code and then automatically creates a patch, derived from the post and instantiated for the developer's code, to resolve the developer's error. 
    \item \textbf{Evaluation.} An evaluation of \toolname on an existing benchmark of \NumBenchmarkSubjects \ac{re} instances, and against 4 baselines.
    \item \textbf{Artifacts.} The public release of our complete dataset, including the \acp{pattern} library and all the patches synthesized and posts found by \toolname and each of the four baselines. \url{https://doi.org/10.6084/m9.figshare.14518407}
\end{itemize}

\section{Illustrative Example}
\label{sec:motivating-example}

\begin{figure*}[t]
\begin{subfigure}[b]{0.49\textwidth}
\begin{subfigure}[b]{\textwidth}
\begin{lstlisting}[language=Java, numberstyle=\scriptsize, rulecolor=\color{black}]
...
listofurls = (String[]) image_urls.toArray(); 
... @\hspace*{\fill}{\small \textbf{Q}}@
\end{lstlisting}
\vspace{-5pt}
\end{subfigure}
\par\vspace{0.2em}
\begin{subfigure}[b]{\textwidth}
\begin{lstlisting}[language=Java, numberstyle=\scriptsize, rulecolor=\color{black}]
listofurls = image_urls.toArray(new String[image_urls.size()]); @\hspace*{\fill}{\small \textbf{A}}@
\end{lstlisting}
\vspace{-7pt}
\caption{\label{fig:so}\acs{so} post \#15264182 question and answer}
\end{subfigure}
\end{subfigure}
\hspace{0.5em}
\begin{subfigure}[b]{0.51\textwidth}
\begin{lstlisting}[language={}, numbers=none, xleftmargin=0em, framexleftmargin=0em]
@\textbf{DELETE}@ class cast @``\textcolor{textRed}{\textbf{(String[])}}"@ on @\textcolor{textRed}{\textbf{line 3}}@
@\textbf{ADD}@ argument @``\textcolor{textRed}{\textbf{new String[...]}}"@ to @``toArray()"@ on @\textcolor{textRed}{\textbf{line 3}}@
@\textbf{ADD}@ argument @``\textcolor{textRed}{\textbf{image\_urls.size()}}"@ to @``new String[...]"@ on @\textcolor{textRed}{\textbf{line 3}}@
\end{lstlisting}
\vspace{-7pt}
\caption{\label{fig:concrete-edit-script}Concrete edit script $\mathbb{S}$ to fix $Q$ using $A$}
\end{subfigure}
\par\vspace{7pt}
\begin{subfigure}[b]{0.48\textwidth}
\begin{lstlisting}[language=Java, numberstyle=\scriptsize, rulecolor=\color{black}, escapeinside={(*}{*)}]
(*@*)Abstract(name="_ABSTRACT_1", val="ArrayList, Set, ...")
public void pattern() {
  _ABSTRACT_1 $v1;
  _WILDCARD_1[] $v2 = (_WILDCARD_1[]) $v1.toArray();
}
\end{lstlisting}
\vspace{-7pt}
\caption{\label{fig:pattern}\ac{pattern} for ClassCastException}
\end{subfigure}
\hspace{0.5em}
\begin{subfigure}[b]{0.52\textwidth}
\begin{subfigure}[b]{\textwidth}
\begin{lstlisting}[language={}, numbers=none, xleftmargin=0em, framexleftmargin=0em, backgroundcolor=\color{textBackgroundGray}]
@\textcolor{textRed}{\textbf{String[]}}@ @$\Longleftrightarrow$@ @\textcolor{textBlue}{\textbf{\_WILDCARD\_1[]}}@, @\textcolor{textRed}{\textbf{image\_urls}}@ @$\Longleftrightarrow$@ @\textcolor{textBlue}{\textbf{\$v1}}@, @\textcolor{textRed}{\textbf{line 3}}@ @$\Longleftrightarrow$@ @\textcolor{textBlue}{\textbf{line 4}}@
\end{lstlisting}
\vspace{-5pt}
\end{subfigure}
\par\vspace{0.15em}
\begin{subfigure}[b]{\textwidth}
\begin{lstlisting}[language={}, numbers=none, xleftmargin=0em, framexleftmargin=0em]
@\textbf{DELETE}@ class cast @``\textcolor{textBlue}{\textbf{(\_WILDCARD\_1[])}}"@ on @\textcolor{textBlue}{\textbf{line 4}}@
@\textbf{ADD}@ argument @``\textcolor{textBlue}{\textbf{new \_WILDCARD\_1[...]}}"@ to @``toArray()"@ on @\textcolor{textBlue}{\textbf{line 4}}@
@\textbf{ADD}@ argument @``\textcolor{textBlue}{\textbf{\$v1.size()}}"@ to @``new \_WILDCARD\_1[...]"@ on @\textcolor{textBlue}{\textbf{line 4}}@
\end{lstlisting}
\vspace{-7pt}
\caption{\label{fig:pattern-edit-script}Generalized edit script $\mathbb{S}^\prime$}
\end{subfigure}
\end{subfigure}
\par\vspace{7pt}
\begin{subfigure}[b]{0.48\textwidth}
\begin{lstlisting}[language=Java, numberstyle=\scriptsize, rulecolor=\color{black}, firstnumber=34]
...
ArrayList<URL> urls;@\lstsetnumber{...}@
...@\lstresetnumber\setcounter{lstnumber}{39}@
@\colorbox{githubRed}{\strut \textbf{-} URL[] array = (URL[]) urls.toArray();}@ @\hspace{0pt} {\small \textbf{\color{red} $\leftarrow$ RE thrown here}}@@\lstresetnumber\setcounter{lstnumber}{39}@
@\colorbox{githubGreen}{\strut \textbf{+} URL[] array = urls.toArray(new URL[urls.size()]);}@@\lstsetnumber{...}@
...@\lstresetnumber\setcounter{lstnumber}{0}@
\end{lstlisting}
\vspace{-7pt}
\caption{\label{fig:buggy}Buggy code and developer's patch (ExtensionService.java)}
\end{subfigure}
\hspace{0.5em}
\begin{subfigure}[b]{0.51\textwidth}
\begin{subfigure}[b]{\textwidth}
\begin{lstlisting}[language={}, numbers=none, xleftmargin=0em, framexleftmargin=0em, backgroundcolor=\color{textBackgroundGray}]
@\textcolor{textBlue}{\textbf{\_WILDCARD\_1[]}}@ @$\Longleftrightarrow$@ @\textcolor{textOrange}{\textbf{URL[]}}@, @\textcolor{textBlue}{\textbf{\$v1}}@ @$\Longleftrightarrow$@ @\textcolor{textOrange}{\textbf{urls}}@, @\textcolor{textBlue}{\textbf{line 4}}@ @$\Longleftrightarrow$@ @\textcolor{textOrange}{\textbf{line 40}}@
\end{lstlisting}
\vspace{-5pt}
\end{subfigure}
\par\vspace{0.15em}
\begin{subfigure}[b]{\textwidth}
\begin{lstlisting}[language={}, numbers=none, xleftmargin=0em, framexleftmargin=0em]
@\textbf{DELETE}@ class cast @``\textcolor{textOrange}{\textbf{(URL[])}}"@ on @\textcolor{textOrange}{\textbf{line 40}}@
@\textbf{ADD}@ argument @``\textcolor{textOrange}{\textbf{new URL[...]}}"@ to ``toArray()" on @\textcolor{textOrange}{\textbf{line 40}}@
@\textbf{ADD}@ argument @``\textcolor{textOrange}{\textbf{urls.size()}}"@ to @``new URL[...]"@ on @\textcolor{textOrange}{\textbf{line 40}}@
\end{lstlisting}
\vspace{-7pt}
\caption{\label{fig:dev-edit-script}Concretized edit script $\mathbb{S}^{\prime\prime}$ to developer's buggy code}
\end{subfigure}
\end{subfigure}
\caption{\label{fig:example}Example from JD-GUI (https://github.com/java-decompiler/jd-gui) throwing ClassCastException}
\vspace{-10pt}
\end{figure*}

We illustrate our technique using the example shown in \Cref{fig:example}. The example is extracted from the JD-GUI project~\cite{jd-gui}, which is a popular interface for viewing Java ``.class" files. The buggy code shown in \Cref{fig:buggy} throws the common \ac{re}, ClassCastException, at \colorbox{githubRed}{\strut line 40} because the \code{toArray()} method returns an \code{Object[]} which cannot be cast to a \code{URL[]}.  

To find a repair for fixing this \ac{re}, the developer may refer to \ac{so}, to find a post discussing the same exception in the same scenario, and then adapt the fix suggested in the answer of the post to their own buggy code. To find such a relevant post, \toolname first searches for a \ac{pattern} best describing the developer's exception scenario (\Cref{fig:pattern}) from our systematically compiled library of \acp{pattern}. Then, this \ac{pattern} is used to find the best \ac{so} post instantiating it (\Cref{fig:so}). 

Consider the \ac{pattern} shown in \Cref{fig:pattern}. It represents an abstract exception-triggering pattern describing the \code{toArray()} failure scenario of ClassCastException. \code{toArray()} is an API defined for the Collections framework, hence the \ac{pattern} abstracts out the concrete data type of variable \code{\$v1}, allowing it to match with any of the Collection classes (\eg \code{ArrayList} and \code{Set}) specified on line 1. Similarly, the specific data type of \code{\$v2} and the casting array is immaterial with reference to the \ac{re}. Therefore, the \ac{pattern} represents both these data types with the same ``\textit{wildcard}" notation encoding an inherent mapping between them. 
The wildcard can be instantiated with any data type, such as \code{URL[]} or \code{String[]}. This \ac{pattern} is identified by \toolname as the best match ($\mathcal{P}_\textit{best}$) as it is perfectly instantiated by the developer's buggy code.

\toolname recommends the \ac{so} post in \Cref{fig:so}, as it perfectly matches $\mathcal{P}_\textit{best}$.
Question code snippets from \ac{so} posts typically include several lines that make the snippets functionally or syntactically complete, but otherwise irrelevant to the exception scenario (shown by $\dots$ in \Cref{fig:so}). This ``noise" may lead to inaccuracies in the patch being extracted from the post, and consequently to an incorrect patch synthesized for the developer's code. To address this problem, \toolname prunes the question code snippet to the relevant lines implicated by $\mathcal{P}_\textit{best}$. \toolname then captures the fix suggested by the answer code snippet in a concrete edit script, $\mathbb{S}$ (\Cref{fig:concrete-edit-script}), by comparing this pruned Q and A code snippets represented by their \acp{apg} -- a simplified and abstracted derivative of the \ac{ast} (see \Cref{sec:apg}). The edit script $\mathbb{S}$ is comprised of a series of three non-trivial changes to the faulty line: (1) \textit{deleting} class cast \code{(String[])}, (2) \textit{adding} a new argument of array instantiation \code{new String[$\dots$]} to \code{toArray()}, and (3) \textit{adding} memory allocation to the newly instantiated array as \code{image\_urls.size()}.

\begin{figure}[t]
  \centering
\begin{subfigure}[b]{0.49\columnwidth}
    \includegraphics[width=\columnwidth]{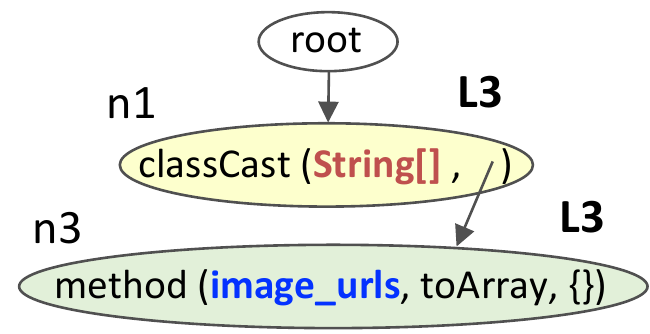}
    \caption{$Q_{\text{APG}}$: \ac{apg} of \Cref{fig:so}}
    \label{fig:so-apg}
\end{subfigure}
\quad
\begin{subfigure}[b]{0.45\columnwidth}
    \includegraphics[width=\columnwidth]{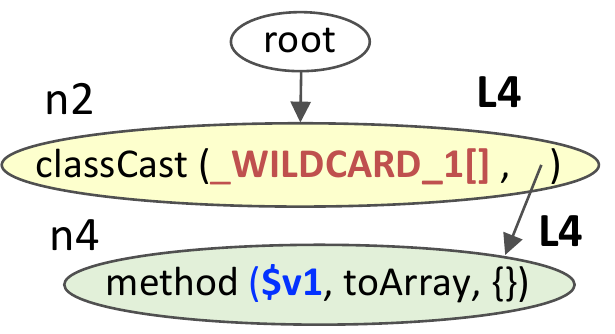}
    \caption{$\mathcal{P}_{\text{APG}}$: \ac{apg} of \Cref{fig:pattern}}
    \label{fig:pattern-apg}
\end{subfigure}
    \caption{\acp{pattern} of \ac{so} question code snippet and \ac{pattern}}
    \label{fig:apgs}
    \vspace{-10pt}
\end{figure}

The next step of \toolname is to abstract out post-specific details from the concrete edit script $\mathbb{S}$ to get a generalized edit script $\mathbb{S}^\prime$. It is challenging to directly generalize the edit script, since points of abstraction vary with different \ac{re} types, as well as with different exception patterns in a specific \ac{re} type. Encoding rules for such abstractions is a cumbersome task. \toolname addresses this challenge by leveraging the abstractions specified in the \ac{pattern}. To establish correspondence, $Q$ and $\mathcal{P}_\textit{best}$ are structurally aligned using their \ac{apg} representations as shown in \Cref{fig:apgs}. The aligned nodes are shown with yellow and green color coding. Further, a fine-grained correspondence between the components of each matched pair of nodes is also established. For example, \code{String[]} from the classCast node in \Cref{fig:so-apg} maps to \code{\_WILDCARD\_1[]} from \Cref{fig:pattern-apg}. Similarly, \code{image\_urls} is matched with \code{\$v1}. This set of mappings as also shown in the gray box of \Cref{fig:pattern-edit-script} are used to adapt $\mathbb{S}$ suitably to produce $\mathbb{S}^\prime$.

Lastly, \toolname concretizes the generalized edit script $\mathbb{S}^\prime$ to the developer's buggy code to get $\mathbb{S}^{\prime\prime}$ as shown in \Cref{fig:dev-edit-script} following the same correspondence and adaptation procedure as described above. As a final step, \toolname applies $\mathbb{S}^{\prime\prime}$ to developer's buggy code to synthesize a repair patch, which is identical to the \colorbox{githubGreen}{\strut developer's patch} shown in \Cref{fig:buggy}. 

The mediation of \acp{pattern} is key to the success of \toolname in generating the correct patch. Without the \ac{pattern}, it would be challenging to determine that the \code{String} data type in the \ac{so} post needs to be migrated to the \code{URL} data type in the developer's code, leading to a incorrect patch. The entire process of finding the right post and generating the correct patch is performed completely automatically by \toolname, using our pre-compiled library of \acp{pattern}.
\section{Approach}
\label{sec:approach}

\begin{figure*}[t]
  \centering
  \includegraphics[width=\textwidth]{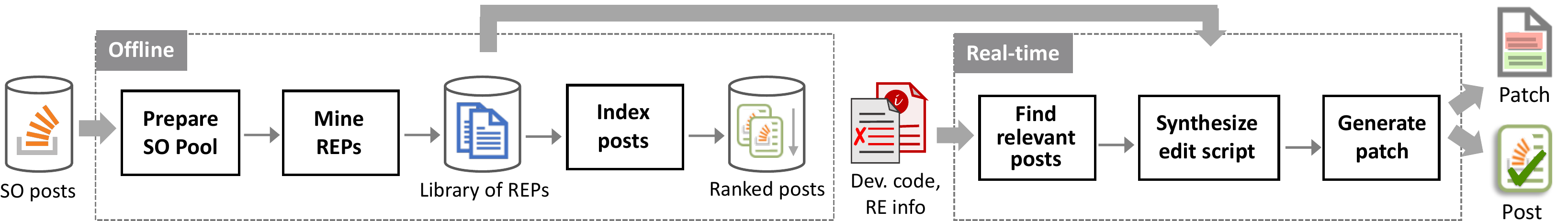}
  \caption{Overview of the approach\label{fig:overview}}
  \vspace{-1em}
\end{figure*}


\toolname's repair strategy is novel and fundamentally different than traditional \ac{apr} techniques in several ways. First, \toolname uses \textit{bug patterns} (\acp{pattern}) to facilitate the generation of repairs, while other \ac{apr} tools (\eg PAR~\cite{kim2013automatic} and Getafix~\cite{GetAFix:OOPSLA:2019}) use a small set of predefined \textit{fix patterns}. Second, indexing \ac{so} with \acp{pattern} allows \toolname to leverage the diverse, but reasonably small space of exemplary concrete repair strategies from \ac{so} posts, while other \ac{apr} tools are required to navigate the huge search space of scenario-agnostic concrete repairs. Lastly, a heavily pruned search space means that \toolname does not require a strong oracle (\eg test cases) to validate patches, allowing \toolname to avoid overfitting and, moreover, achieve a real-time performance.

Our approach is designed around a library of \acp{pattern} (\textit{bug patterns}), semi-automatically mined from \ac{so} posts. Specifically, a \ac{pattern} is an abstracted sequence of Java statements representing an exception-triggering scenario. The \acp{pattern} play a central role in mediating all aspects of \toolname's operation, from ranking and searching of \ac{so} posts, to synthesizing patches from a given post to fixing an \ac{re} in a developer's code. Compiling the \ac{pattern} library is a \textit{low-cost}, \textit{incremental}, \textit{one-time effort}. Low-cost because only basic Java knowledge is required to read a post and write its \ac{pattern}. One-time because once the \acp{pattern} are written, their benefit can be harvested thereafter without any extra cost. Incremental because new \acp{pattern} can be gradually added to the library for new \ac{so} posts.

\Cref{fig:overview} shows an overview of \toolname's repair approach, consisting of four main stages: \textit{preparing} the \ac{so} pool for analysis, \textit{mining} \acp{pattern} from \ac{so} posts, \textit{indexing} of \ac{so} posts, and \textit{fixing} \acp{re} in a developer's buggy code. The preparing, mining, and indexing stages are performed offline, while the fixing stage is performed in real-time.


\subsection{\acf{apg}}
\label{sec:apg}

We use the \ac{apg} representation proposed in our prior work~\cite{MAESTRO:FSE2020} for analyzing code snippets.
Briefly, the \ac{apg} is a simplified and abstracted derivative of the \ac{ast}. It captures the structural relationships between program statements, while normalizing low-level syntactic details to facilitate a meaningful comparison between code snippets with significant differences in variable names, data types, and program constructs (\eg while vs. for loop). For example, \Cref{fig:so-apg} shows the  \ac{apg} for the code snippet in \Cref{fig:so}. The similarity score between two \acp{apg} is computed by aligning them using the APTED tree edit distance algorithm~\cite{APTED2015,APTED2016}, and cumulatively counting the normalized number of matching components in a pair of corresponding nodes. For example, in \Cref{fig:apgs}, the similarity score is 1.0 as the component  \code{String[]} matches with the wildcard in $\langle n1, n2\rangle$. Similarly, the components, type of method caller (\code{ArrayList}) and method name (\code{toArray}), in $\langle n3, n4\rangle$ match perfectly.

\subsection{Stage 1: Preparing the \acf{so} pool}
\label{sec:so-pool-preparation}

In this stage, \toolname selects \ac{re}-related posts and groups them by \ac{re} type. An \ac{so} post is selected if it has: (1) \ac{re} type in the title, (2) ``java" or ``android" tags, (3) at least one answer, and (4) at least one \emph{parsable} question code snippet. 

A snippet is considered parsable if it is syntactically well-formed and 
can be parsed using any off-the-shelf Java parser (\eg Eclipse JDT~\cite{EclipseJDT}). However, a large number of code snippets in \ac{so} are found to be malformed with missing or extraneous syntactic characters (\eg parentheses) and/or undefined tokens (\eg an ellipsis (...) in place of actual code), rendering them unusable by our approach~\cite{Yang:MSR2016, CSNIPPEX:ISSTA2016}.

Addressing this concern, we propose an error-driven iterative approach to 
automatically repair unparsable snippets to make them parsable.
For a given code snippet, our approach picks a parsing error $e \in \mathbb{E}$, applies a fix for $e$, and continues this process as long as the number of parsing errors are reducing ($|\mathbb{E}| < | \mathbb{E}_{\textit{prev}}|$) or the snippet is now parsable ($\mathbb{E} = \emptyset$). We use Eclipse JDT for populating $\mathbb{E}$. For an error $e$, defined by the problematic element $\mathbb{T}$ at location $\mathcal{L}$, our approach applies a fix based on the rules shown in \Cref{tab:unparsable-to-parsable}. For example, for the invalid token \code{"..."} in snippet \code{List x = ...;}, our approach corrects it with the fixing rule of applying a suitable initialization expression: \code{List x = new ArrayList<>()}. Similarly, other examples include inserting a missing semi-colon (\code{;}) at the end of a statement, inserting a missing \code{catch} block after a \code{try}, or removing an extra brace (\code{\}}) from the end of a method declaration.


\begin{table}[t]\scriptsize
\centering
{\renewcommand{\arraystretch}{1.2}%
  \begin{tabular}{|l|l|}
    \hline
    \textbf{Error} & \textbf{Fixing rule}\\
    \hline
    Invalid token $\mathbb{T}$ at $\mathcal{L}$ & $\mathcal{L}$ $\in$ Expression $\longrightarrow$ replace $\mathbb{T}$ with $\mathbb{T}_{\textit{valid}}$ at $\mathcal{L}$\\
    & $\mathcal{L}$ $=$ Statement $\longrightarrow$ delete $\mathbb{T}$ at $\mathcal{L}$\\
    \hline
    Missing $\mathbb{T}$ at $\mathcal{L}$ & $\mathbb{T}$ $\in$ Symbol $\longrightarrow$ insert $\mathbb{T}$ at $\mathcal{L}$\\
    & $\mathbb{T}$ $\in$ Grammar Rule $\longrightarrow$ insert instantiation of  $\mathbb{T}$ at $\mathcal{L}$\\
    \hline
    Extra token $\mathbb{T}$ at $\mathcal{L}$ & delete $\mathbb{T}$ at $\mathcal{L}$\\
    \hline
  \end{tabular}
  }
  \vspace{-3pt}
  \caption{Rules for fixing parsing errors in code snippets}
  \label{tab:unparsable-to-parsable}
  \vspace{-7pt}
\end{table}
\subsection{Representation for \acfp{pattern}}
\label{sec:pattern}


\setlength{\grammarparsep}{1px} 
\grammarindent=55pt
\AtBeginEnvironment{grammar}{\footnotesize}
\begin{figure}
\begin{mdframed}[innerleftmargin=3pt, innerrightmargin=8pt, innertopmargin=2pt, innerbottommargin=2pt]
\begin{grammar}

<pattern> ::= (<Java statement>)+

<identifier> ::= <Java identifier> \alt
'{\_ABSTRACT\_}'[1-9]+ | '{\_WILDCARD\_}'[1-9]+

<annotation> ::= '@Abstract (name={\_ABSTRACT\_}' [1-9]+ ',' 'val=' \{set of permissible values\} ')'

\end{grammar}
\end{mdframed}
\vspace{-10pt}
\caption{\label{fig:grammar} Extensions to Java grammar for writing \ac{pattern}}
\vspace{-1.5em}
\end{figure}

Our \ac{pattern} conforms to Java grammar rules, with a couple of straightforward extensions to support generalization, shown in \Cref{fig:grammar}. We define two types of generalizations in representing identifiers, such as method names and data types. The first is \textit{wildcard}, to represent zero or more program elements. For example, the data type of the casting array in \Cref{fig:pattern} is specified with a wildcard since it could be instantiated with any of the Java or user-defined types, such as \code{String[]} or \code{URL[]}. The second generalization is called \textit{abstract semantics}, which allows specifying a set of permissible semantic equivalent values using Java annotations. For example, the \ac{pattern} shown in \Cref{fig:pattern} represents the data type of ``\$v1" as a semantic abstraction, since it can only be instantiated with a Collection class, such as \code{List}, \code{ArrayList}, or \code{Set}.

\subsection{Stage 2: Mining \acfp{pattern}} 
\label{sec:mining}

\begin{figure}[t]
  \centering
  \includegraphics[width=0.95\columnwidth]{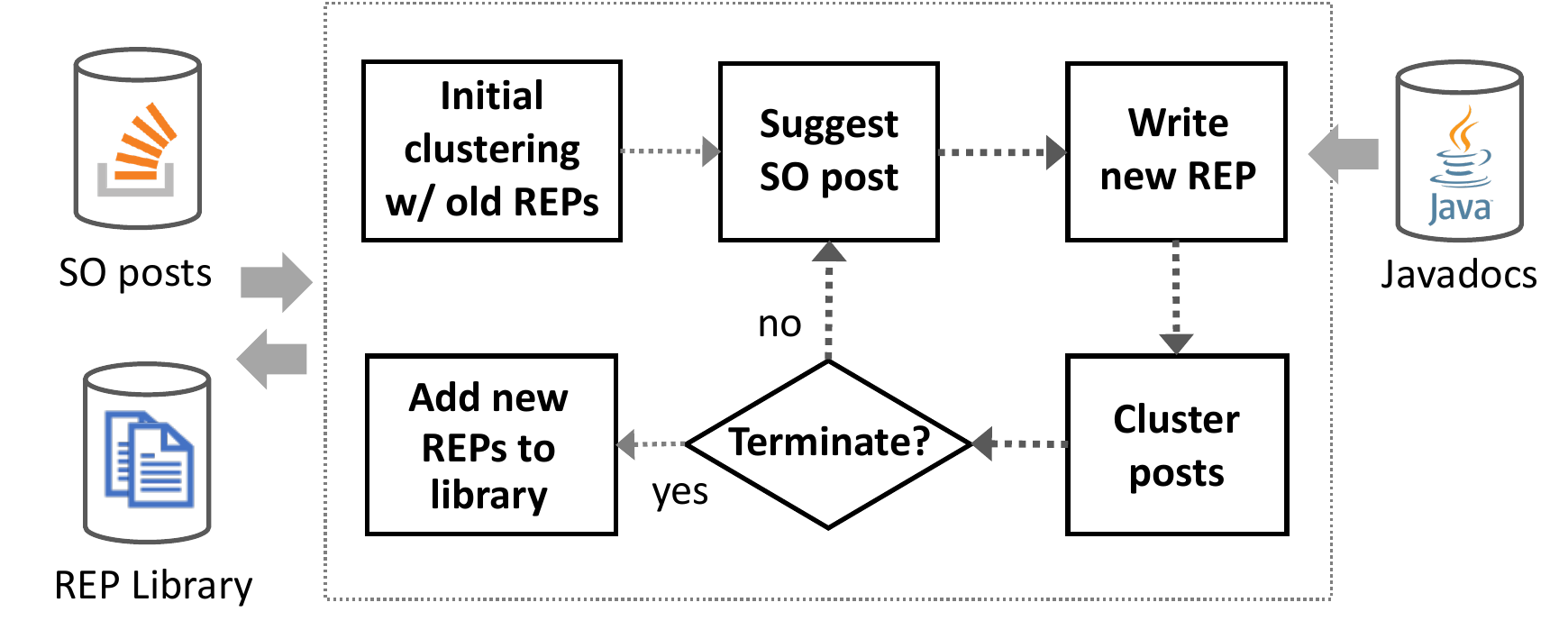}
  \vspace{-3pt}
  \caption{Overview of the \ac{pattern} mining process\label{fig:mining}}
  \vspace{-1.5em}
\end{figure}

Our approach is predicated on \emph{precisely} extracting the exception-triggering bug patterns, \ie \acp{pattern} represented in \ac{so} posts. Our previous work~\cite{MAESTRO:FSE2020} highlighted the challenges of doing so fully automatically, particularly because of the poor quality of Q\&A snippets. On the other hand manually writing \acp{pattern} for each of thousands of \ac{so} posts is impractical and also redundant, since several \ac{so} posts exhibit the same \ac{pattern}. Therefore, we adopt a semi-automated approach, shown in \Cref{fig:mining}, that optimally combines human knowledge and automation. Specifically, we iteratively ask the human to \emph{manually} author a \ac{pattern} based on a single \emph{representative} \ac{so} post and then \emph{automatically} cluster all other posts instantiating the same \ac{pattern}, removing them from the pool of human inspection. Key features of our mining algorithm include \textit{inexpensive}, \textit{one-time} effort of writing \acp{pattern}, with support for \textit{incrementally} expanding the library of \acp{pattern} for new \ac{so} posts.

\subsubsection{\toolname suggests an \ac{so} post} 
In this step \toolname identifies an unvisited \ac{so} post that exemplifies a commonly occurring \ac{pattern}. To this end, \toolname picks the top post after ordering unvisited \ac{so} posts (for an \ac{re} type) by user votes, breaking ties using mean distance of the candidate post from the current pool of \acp{pattern}. The distance is the inverse of similarity score (\Cref{sec:apg}) between a \ac{pattern} and a post's question code snippet.

\begin{figure*}[t]
\centering
\begin{subfigure}[b]{0.34\textwidth}
    \fbox{\includegraphics[width=\textwidth]{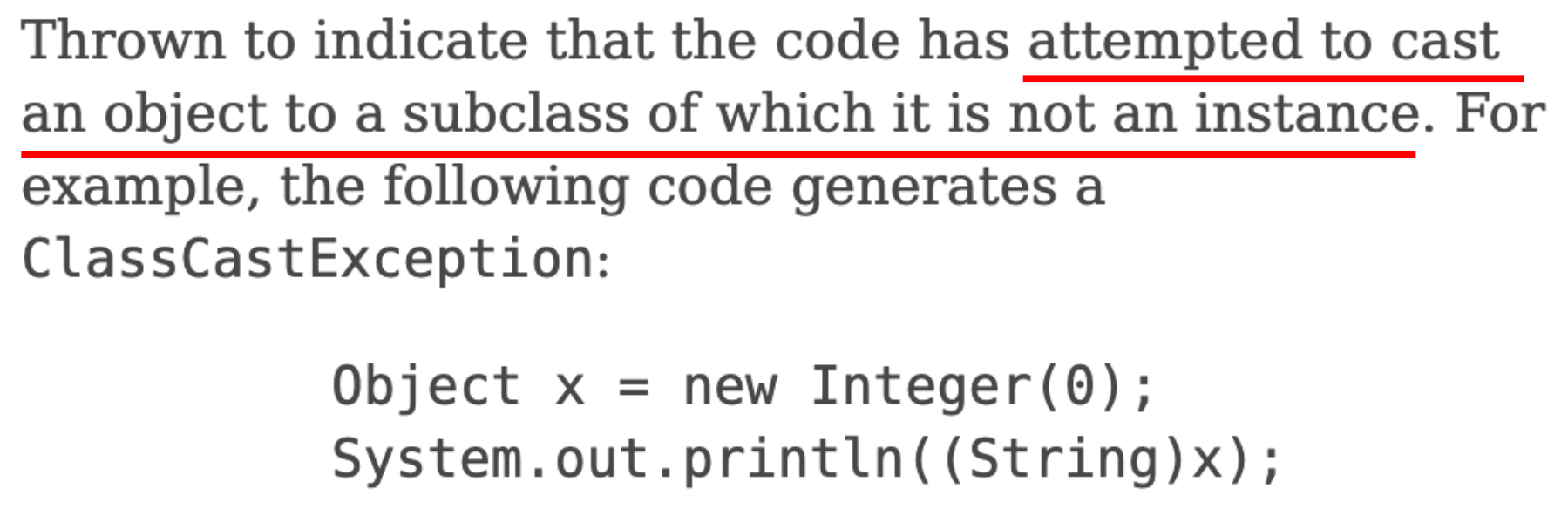}}
    \caption{Explanation of ClassCastException}
    \label{fig:mining-eg-1}
\end{subfigure}
\hspace{0.1em}
\begin{subfigure}[b]{0.34\textwidth}
    \fbox{\includegraphics[width=\textwidth]{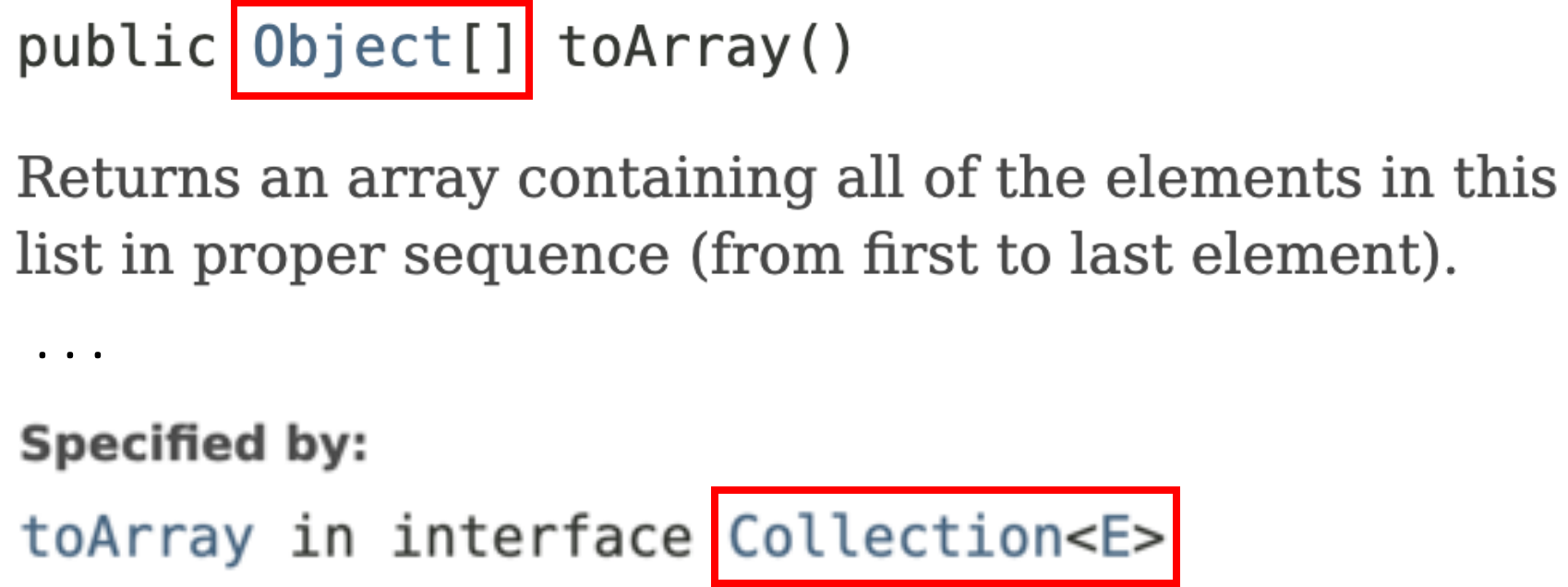}}
    \caption{ArrayList Javadoc for \code{toArray()} API}
    \label{fig:mining-eg-2}
\end{subfigure}
\hspace{0.1em}
\begin{subfigure}[b]{0.29\textwidth}
    \fbox{\includegraphics[width=\textwidth]{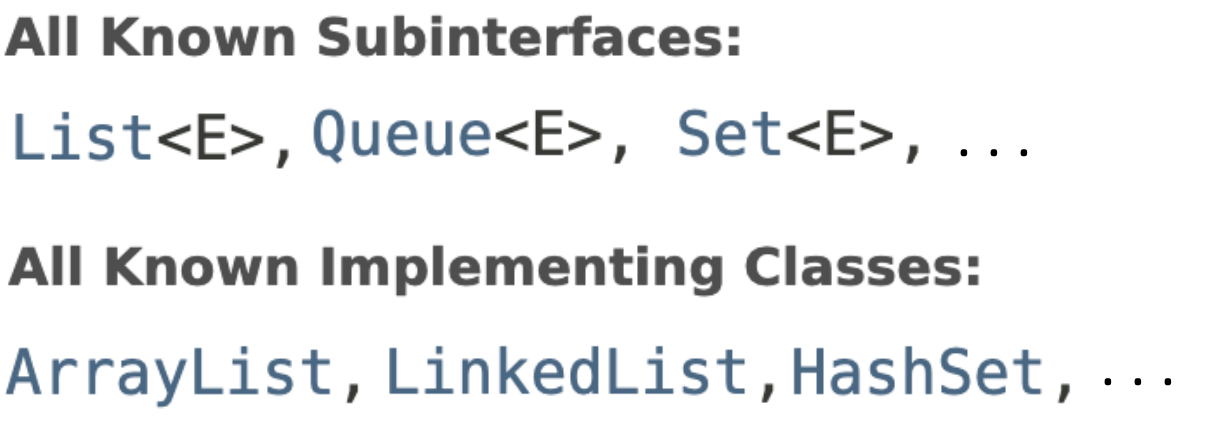}}
    \caption{Specification of \code{Collection$<$E$>$}}
    \label{fig:mining-eg-3}
\end{subfigure}
\vspace{-15pt}
\caption{\label{fig:mining-eg}JavaDocs helpful in writing the \ac{pattern} shown in \Cref{fig:pattern} from \ac{so} post shown in \Cref{fig:so}}
\vspace{-5pt}
\end{figure*}

\subsubsection{Human writes a \ac{pattern}} 
This step involves reviewing two specific sources, namely the suggested \ac{so} post and relevant Javadocs, to write the \ac{pattern}. We expect that following the below procedure a person reasonably experienced in Java, need not be an expert, should be able to write the \acp{pattern}. First, the human studies the suggested \ac{so} post, particularly focusing on the question text and code snippet, to extract the post-specific exception scenario. To assist in this process, our approach highlights the potential relevant lines in the question code snippets that are referenced in the answer. To further generalize the post-specific exception scenario, the human may refer to the official Java documentation~\cite{javadocs} of the \ac{re} as well as the different classes or APIs used in the post code. 

For example, consider the \ac{so} post shown in \Cref{fig:so}. We now explain how the human can write the \ac{pattern} shown in \Cref{fig:pattern} from this post. First, the human can capture the post-specific scenario as shown in line 2. To abstract out the post-specific details, the human could then refer to the Javadocs shown in \Cref{fig:mining-eg}. For instance, the human could visit the definition of ClassCastException (\Cref{fig:mining-eg-1}) to know that casting to \code{String[]} in our example is the problem. Then, the human could explore the Javadocs of \code{ArrayList}, which is the data type of the \code{image\_urls} variable, specifically focusing on the \code{toArray()} API (\Cref{fig:mining-eg-2}). Upon studying this documentation, the human can figure out two generalizations: (1) \code{Object[]} returned by the \code{toArray()} API when cast to any other array type would cause the \ac{re}, thus, the concrete casting type \code{String[]} could be safely generalized to a \textit{wildcard}, and (2) \code{toArray()} API is common for all implementations of the \code{Collection} framework (highlighted in \Cref{fig:mining-eg-2}), thus, the data type of the \code{image\_urls} variable could be represented using \textit{abstract semantics}. The set of values for this could be extracted from the Javadoc of the \code{Collection} framework (\Cref{fig:mining-eg-3}), which comprehensively lists all concrete instantiations (\eg List, Set, and ArrayList). Finally, the human could normalize the \ac{pattern} by renaming the variables to generic names (\eg\xspace\code{image\_urls} to \code{\$v1}) to get the generalized \ac{pattern} shown in \Cref{fig:pattern}.

Alternatively, the human can simply mark the suggested post as \emph{unviable} if a unique or meaningful \ac{pattern} cannot be extracted from the it. Another post is suggested in this case.

\subsubsection{\toolname clusters posts by \ac{pattern}}
The goal is to group together \ac{so} posts that constitute instances of a newly authored \ac{pattern}, \ie question code snippets of posts having a perfect similarity score (1.0) with the \ac{pattern}. (Similarity score is computed as described in \Cref{sec:apg}.) Clustered posts are then marked as visited. This smartly eliminates redundancies in the work performed by the human. For \textit{incremental} compilation of the \ac{pattern} library, an \textit{initial clustering} is performed to group new \ac{so} posts instantiating existing \ac{pattern} library, so that the human is suggested only truly ``unseen" posts.

\subsubsection{Termination condition}
The mining process terminates if all posts are exhausted or when $U$ consecutive posts are marked as unviable by the human. The latter indicates a point of diminishing returns, conceivably because most distinct, popular \acp{pattern} have already been extracted.


\subsection{Stage 3: Indexing of \ac{so} Posts} 
\label{sec:indexing}

Analyzing an \ac{so} post to generate a repair patch is computationally expensive. Hence, effective indexing of \ac{so} posts plays an important role in the success of our approach in achieving the real-time use case. 
For each cluster of \ac{so} posts from the mining step (\Cref{sec:mining}), our indexing ranks the posts based on their relevance to the cluster's representative \ac{pattern}. Particularly, we compute the similarity score between a \ac{pattern} and each answer code snippet in a post to up-rank posts with answers that suggest a complete fix with regards to the \ac{pattern}, while down-ranking posts with partial or irrelevant answers. Thus, for our illustrative example (\Cref{fig:example}), post \#1524182 gets a high rank since it matches closely to the \ac{pattern}. Post \#16656384 gets a lower rank since it gives only a partial answer with missing assignment operation. Post \#46201465 gets an even lower rank, since its answer is completely irrelevant to the \ac{pattern}. For posts with multiple answers, the highest scoring answer is used for the ranking.



\subsection{Stage 4: Fixing \ac{re} in Developer's Code}
\label{sec:fixing}


\setlength{\textfloatsep}{0.1cm}
\setlength{\floatsep}{0.1cm}
\begin{algorithm}[t]\scriptsize
\caption{Algorithm for Fixing \ac{re} in Developer's Code}
\label{alg:fixing}
\textbf{Input:} \hspace*{0.5em} $\mathcal{B}$: Developer's buggy code\\
\hspace*{3.5em} $E$: Exception information (\ac{re} type and failing line number)\\
\hspace*{3.5em} $\mathcal{P} = $\{$p_1, ..., p_n$\}:  Library of \acp{pattern}\\
\hspace*{3.5em} $\mathcal{I} = $\{$s_1, ..., s_n$\}:  Library of indexed \ac{so} posts and their Q\&A pairs\\
\hspace*{3.5em} $\mathcal{K}$:  Number of repair patches to generate\\
\hspace*{3.5em} $\mathcal{Z}$:  Number of \ac{so} posts to evaluate\\
\textbf{Output:} $\mathcal{O} = $\{$\langle r_1, s_1\rangle,...,\langle r_k, s_k\rangle$\}: List of $k$ patch-post pairs\\
$\textbf{begin}$
\begin{algorithmic}[1]

\State $\mathcal{O}$ $\gets$ \{\}
\State \highlightgray{/* Step 1. Find relevant \ac{so} posts */}
\State $\mathcal{P}_{\mathit{best}}$ $\gets$ findBestPattern($\mathcal{P}$, $\mathcal{B}$, $E$)
\State $\mathcal{S}$ $\gets$ getRankedPosts($\mathcal{I}$, $\mathcal{P}_{\mathit{best}}$)
\State $P_{\text{APG}}$ $\gets$ buildAPG($\mathcal{P}_{\text{best}}$)

\State postCount $\gets$ 0
\For {\textbf{each} $s$ $\in$ $\mathcal{S}$}
    \State postCount $\gets$ postCount + 1
    \State $\mathit{qaPairs}$ $\gets$ getQAPairs($\mathcal{I}$, $s$)
    \For {\textbf{each} $\langle Q, A\rangle$ $\in$ $\mathit{qaPairs}$}
        \State \highlightgray{/* Step 2. Clean Q and A code snippets */}
        \State $Q_{\text{APG}}$ $\gets$ buildAPG($Q$)
        \State $A_{\text{APG}}$ $\gets$ buildAPG($A$)
        \State $T$ $\gets$ triangulate($Q_{\text{APG}}$, $A_{\text{APG}}$, $P_{\text{APG}}$)
        \State $Q_{\text{APG}}$ $\gets$ pruneAPG($Q_{\text{APG}}$, $T.$quesRelevantLines)
        \State $A_{\text{APG}}$ $\gets$ pruneAPG($A_{\text{APG}}$, $T.$ansRelevantLines)
        
        \State \highlightgray{/* Step 3. Synthesize Generalized Edit Script */}
        \State $\mathbb{S}$ $\gets$ getEditScript($Q_{\text{APG}}$, $A_{\text{APG}}$)
        \State $\mathbb{S}^\prime$ $\gets$ adaptEditScript($\mathbb{S}$, $Q_{\text{APG}}$, $P_{\text{APG}}$)
        
        \State \highlightgray{/* Step 4. Generate repair patch */}
        \State $\mathbb{S}^{\prime\prime}$ $\gets$ adaptEditScript($\mathbb{S}^\prime$, $P_{\text{APG}}$, $\mathcal{B}_{\text{APG}}$)
        \State $\mathcal{B}^\prime_{\text{APG}}$ $\gets$ applyChanges($\mathcal{B}_{\text{APG}}$, $\mathbb{S}^{\prime\prime}$)
        \If {isValid($\mathcal{B}^\prime_{\text{APG}}$) = \textit{true}}
            \State $r$ $\gets$ convertAPGToJavaPatch($\mathcal{B}^\prime_{\text{APG}}$)
            \If {isParsable($r$) = \textit{true}}
                \State $\mathcal{O}$ $\gets$ $\mathcal{O}$ $\cup$ $\langle r, s\rangle$
            \EndIf
        \EndIf
        \State \highlightgray{/* Step 5. Check termination criteria */}
        \If {$|\mathcal{O}|$ $\geq$ $\mathcal{K}$ \textbf{or} postCount $\geq$ $\mathcal{Z}$}
            \If {$|\mathcal{O}|$ = 0}
                \State $\mathcal{O}$ $\gets$ $\langle null, \mathcal{S}[1]\rangle$
            \EndIf
            \State \Return $\mathcal{O}$
        \EndIf
    \EndFor
\EndFor
\State \Return $\mathcal{O}$

\end{algorithmic}
\textbf{end}
\end{algorithm}

\begin{algorithm}[t]\scriptsize
\caption{Algorithm for Adapting Edit Script}
\label{alg:fixing-adapt}
\textbf{Input:} \hspace*{0.5em} $\mathcal{T}$: Edit script to be adapted\\
\hspace*{3.5em} $X$: \ac{apg} w.r.t. $\mathcal{T}$ (source \ac{apg})\\
\hspace*{3.5em} $Y$: \ac{apg} to be used for adaptation (target \ac{apg})\\
\textbf{Output:} $\mathcal{T}^\prime$: Adapted edit script\\
$\textbf{begin}$
\begin{algorithmic}[1]

\State $\mathcal{T}^\prime$ $\gets$ \{\}
\State $\mathcal{N}$ $\gets$ computeMatchedNodes($X$, $Y$)
\State $\mathcal{C}$ $\gets$ computeCorrespondingComponents($\mathcal{N}$)
\For {\textbf{each} $\mathit{op}$ $\in$ $\mathcal{T}$}
    \State $n^\prime$ $\gets$ getMatchedNode($\mathcal{N}$, $\mathit{op}$.$n$)
    \If {$\mathit{op}$.$\mathit{type}$ = \texttt{add} \textbf{or} $\mathit{op}$.$\mathit{type}$ = \texttt{update} \textbf{or} $\mathit{op}$.$\mathit{type}$ = \texttt{replace}}
        \State $m^\prime$ $\gets$ updateCorrespondingComponents($\mathit{op}$.$m$, $\mathcal{C}$)
    \EndIf
    \State $\mathit{op}^\prime$ $\gets$ buildEditOperation($\mathit{op}$.$\mathit{type}$, $n^\prime$, $m^\prime$, $\mathit{op}$.$\mathit{pos}$)
    \If {isValid($\mathit{op}^\prime$, $Y$) = \textit{true}}
        \State $\mathcal{T}^\prime$ $\gets$ $\mathcal{T}^\prime$ $\cup$ $\mathit{op}^\prime$
    \EndIf
    \State \Return $\mathcal{T}^\prime$
\EndFor
\end{algorithmic}
\textbf{end}
\end{algorithm}

\Cref{alg:fixing} shows the overall algorithm. Three sets of input are required. First is from the developer's failure: buggy code, $\mathcal{B}$, and \ac{re} information, $E$. Second is from \toolname's offline analysis: the library of mined \acp{pattern}, $\mathcal{P}$, and the library of indexed \ac{so} posts and their Q\&A pairs, $\mathcal{I}$. Third is a set of configurable parameters: number of patches to generate, $\mathcal{K}$, and number of \ac{so} posts to analyze, $\mathcal{Z}$. The output of the approach is a list $\mathcal{O}$ of $\langle r, s\rangle$ pairs, where $r$ is a repair patch synthesized from an \ac{so} post $s$.

\textbf{Step 1. Find Relevant \ac{so} Posts.} The initial part of the algorithm (lines 3 and 4) finds the \ac{pattern}, $\mathcal{P_{\mathit{best}}}$, best matching the \ac{re}-throwing developer code, and fetches relevant \ac{so} posts for $\mathcal{P_{\mathit{best}}}$ from the indexed library. The algorithm then iterates over each \ac{so} post and its Q\&A pairs to perform steps 2--5.

\textbf{Step 2. Clean Q and A code snippets.} Q and A  snippets may include code to make the snippet functionally or syntactically complete, but otherwise irrelevant to the exception scenario and its repair. If used as such, the repair patch extracted from such snippets would likely be noisy and incorrect. Therefore, the second step of the algorithm (lines 12--16) aims to prune the snippets to relevant lines by triangulating $Q$, $A$, and $\mathcal{P_{\mathit{best}}}$, and retaining any newly inserted fix lines (\eg null check for NullPointerException).

\textbf{Step 3. Synthesize Generalized Edit Script.} The goal is to derive a script, $\mathbb{S}^\prime$, that is post-agnostic. We use the APTED~\cite{APTED2015, APTED2016} tree-edit distance algorithm to first compute the concrete edit script, $\mathbb{S}$ (line 18), which suggests edit operations for fixing the \ac{re} in $Q$ as prescribed by $A$. Each entry in our edit script is comprised of one of the following four edit operations:
\begin{itemize}
    \item \code{add($n$,$m$,$\mathit{pos}$)}: Insert a new node $m$ in the \ac{apg} at position $\mathit{pos}$ (parent or child) with reference to node $n$
    \item \code{delete($n$)}: Delete node $n$ from the \ac{apg}
    \item \code{update($n$,$m$)}: Update value of $n$ with the value of $m$
    \item \code{replace($n$,$m$)}: Replace the subtree rooted at node $n$ with the subtree given by new node $m$
\end{itemize}

For generalizing $\mathbb{S}$ (line 19), our insight is to adapt it in the context of $\mathcal{P_{\mathit{best}}}$, since, by definition, \acp{pattern} exemplify an \textit{abstracted} description of the exception-raising scenario. We use the adaptation algorithm shown in \Cref{alg:fixing-adapt} (and discussed below) to obtain the generalized edit script, $\mathbb{S}^\prime$.

\textbf{Adapting edit script}. \Cref{alg:fixing-adapt} takes as input the edit script to be adapted, $\mathcal{T}$, and the source and target \acp{apg}, $X$ and $Y$, to output the adapted edit script, $\mathcal{T}^\prime$. The algorithm begins by computing a function $M$ : $X \mapsto Y$ that aligns $X$ and $Y$ to get pairs of matching nodes ($\mathcal{N}$) and pairs of corresponding components ($\mathcal{C}$) for $\mathcal{N}$ (lines 2 and 3). Then for each edit operation, $\mathit{op}$ $\in$ $\mathcal{T}$, the algorithm uses this equivalence information to suitably adapt $\mathit{op}$ (lines 5--8). First, the anchor node $n$ is updated with the matched node $n^\prime$. Next, components, such as variable names, types, and method names, in node $m$ are updated with corresponding values from $Y$ to get $m^\prime$. The adapted edit operation, $\mathit{op}^\prime$, is then checked for correctness in the context of $Y$ with two checks: (1) $n^\prime$ is not empty and (2) identifiers in $m^\prime$ are from the namespace of $Y$ (line 10). If the operation $\mathit{op}^\prime$ is found to pass these two checks, then it is added to the list of adapted script $\mathcal{T}^\prime$.

\textbf{Step 4. Generate Repair Patch.}
The goal of this step is to concretize the generalized edit script to developer's buggy code and generate a repair patch (lines 21--28). First, the same modular ``adaptEditScript()" algorithm (\Cref{alg:fixing-adapt}) is used to obtain the concretized script, $\mathbb{S}^{\prime\prime}$. Then, the fixes (edit operations) in $\mathbb{S}^{\prime\prime}$ are applied to developer's buggy code \ac{apg} to get a modified version, $\mathcal{B}^\prime_{\text{APG}}$. This \ac{apg} is then validated for well-formedness with checks, such as no cycles. The $\mathcal{B}^\prime_{\text{APG}}$ is then translated to a Java repair patch $r$. The patch $r$ is then validated for parsability by running it through a Java parser, such as Eclipse JDT (line 25). If parsable, $r$ is then added to the output list along with the \ac{so} post, $s$.

\textbf{Step 5. Check Termination Criteria.}
The algorithm terminates if: (1) $\mathcal{K}$ repair patches are generated or (2) $\mathcal{Z}$ \ac{so} posts have been analyzed. Upon termination, output $\mathcal{O}$ is returned with up to $\mathcal{K}$ patch-post pairs, or only the top-post if a patch could not be generated (lines 30-35).

\begin{table}[t]\scriptsize
  \centering
  \setcellgapes{2pt}\makegapedcells
  \begin{tabular}{|@{\hskip 3pt}p{1.45cm}|p{6.5cm}|}
    \hline
    \multicolumn{2}{|c|}{\textbf{Rating scale for judging repair patches}}\\ 
    \hline
    \textit{Correct} & Patch is identical or semantically equivalent to developer patch\\
    \hline
    \textit{Almost correct} & Patch has one-token difference from correct patch\\
   \hline
    \textit{No patch} & Patch could not be synthesized by the tool\\
    \hline
    \textit{Incorrect} & Patch is incapable of fixing the \ac{re}\\
    \hline
    \hline
    \multicolumn{2}{|c|}{\textbf{Rating scale for judging \ac{so} posts}}\\ 
    \hline
    \textit{Perfect} & Post suggests an accurate repair for the \ac{re} scenario \\
    \hline
    \textit{Helpful} & Post is informative, but no direct repair offered \\
   \hline
    \textit{No post} & Post was not recommended by the tool \\
    \hline
    \textit{Irrelevant} & Post is misleading for repairing the \ac{re} \\
    \hline
  \end{tabular}
  \caption{Rating scales for patches and posts}
  \label{tab:rating}
\end{table}

\section{Evaluation}
\label{sec:evaluation}
Our evaluation addresses the following research questions:

\RQ{RQ1}{How effective is \toolname in assisting fixing of \acp{re}?}

\RQ{RQ2}{How does \toolname perform against other techniques?}

\RQ{RQ3}{How effective are the key contributions in \toolname?}

\RQ{RQ4}{What is the cost of maintaining the \ac{pattern} library?}

\subsection{Implementation}
We implemented our approach in Java as a prototype tool named \toolname (\textbf{M}ine and \textbf{A}nalyz\textbf{E} \textbf{ST}ackoverflow to fix \textbf{R}untime excepti\textbf{O}ns). We used Eclipse JDT~\cite{EclipseJDT} to verify parsability of code snippets and build them into \acp{ast}. We used the APTED tree edit distance algorithm~\cite{APTED2015, APTED2016, APTEDImpl} to compute the preliminary Q\&A edit script, which we augmented with the ``update" and ``replace" edit operations. For the \ac{pattern} \textit{mining} process, we empirically selected the  termination criteria value as $U$ = 3 (\Cref{sec:mining}) to stop the algorithm when it starts to repeatedly suggest low-quality \ac{so} posts. For the \textit{fixing} algorithm discussed in \Cref{sec:fixing}, we set the following values. The number of \ac{so} posts to be evaluated is set as $\mathcal{Z}$ = 15. We chose this value with the real-time use case in mind, since analyzing an \ac{so} post for repair is an expensive operation. The value for the number of patches to generate is set to $\mathcal{K}$ = 3 with the rationale that developers often only inspect a few top patches from a ranked list~\cite{parnin:ISSTA2011}.

\subsection{Datasets}
\textbf{\ac{so} Pool.} 
We used the \ac{so} data dump released in released in March 2019~\cite{SODump}. Based on the selection criteria discussed in \Cref{sec:so-pool-preparation}, this gave us a pool of 24,343 usable \ac{so} posts. The number of posts per \ac{re} type ranged from 3 to 13,415, with an average of 1281 posts and a median of 128 posts. Out of the total 115,009 code snippets evaluated, about 60\% were found to be readily parsable. With our algorithm of converting unparsable snippets to parsable, we were able harvest a significant 16,799 more snippets, thereby improving the overall number of parsable snippets to 75\%. 

\textbf{Library of \acp{pattern}.}
Our \ac{pattern} library is comprised of 158 \acp{pattern}, clustering 10,143 posts. \toolname, on average, suggested 17 \ac{so} posts per \ac{re} type to the human for processing, \acp{pattern} were written for roughly half and the other half were marked as unviable. On average, 8 \acp{pattern} per \ac{re} type resulted in the clustering of 533 posts, implying that the human is required to write only one \ac{pattern} per 67 \ac{so} posts. From the perspective of \ac{so} pool, writing \acp{pattern} for 0.75\% of the \ac{so} posts gives a coverage of over 50\%, on average. This indicates that a small amount of human effort can yield large number of \acp{pattern}.

\textbf{Benchmark.}
For our experiments, we use our publicly released benchmark~\cite{MAESTRO:FSE2020, maestro-repo}. It is an \ac{re}-specific benchmark, comprising 78 instances spanning 19 \ac{re} types collected from the top-500 Java projects on GitHub. 

Defects4J~\cite{defects4j:ISSTA2014, defects4j} is a popular benchmark used by state of the art \ac{apr} techniques. However, it is not suitable for our use case for two reasons. First, Defects4J is a dataset for general-purpose bugs, with only a few instances (less than 10\%) related to \ac{re} failures. Furthermore, many of these instances also require non-\ac{re} fixes to completely resolve the bug, which is out of scope for our technique. Second, the usable instances from Defects4J are limited in diversity, covering only a few common \ac{re} types, such as NullPointerException. 

\subsection{Evaluation Methodology}
\label{sec:protocol}
We evaluate \toolname and its baselines in terms of the number of useful artifacts that they produce, \ie the ability to recommend a relevant \ac{so} post and correctness of the synthesized patch. We use manual examination for patch validation, which is a recommended protocol in the \ac{apr} community~\cite{cocount:ISSTA2020, Phoenix:FSE2019, elixir2017, Angelix:ICSE2016, Kali:ISSTA2015, Durieux:CoRR:2015, SPR:FSE2015, APREvaluation:JSS2021}.

\textbf{Participants and Protocol.}
To avoid any bias in the evaluation, we recruited two external participants to judge the artifacts. Our participants are software professionals with over 10 years of Java experience. For each of the five tools (\toolname and 4 baselines), the participants were presented with up to 3 patch-post pairs for each instance and asked to provide a rating for the best pair based on the metrics discussed below. To reduce bias in the experiment, we presented the results in randomized order with the tool names anonymized. The participants rated all of the 390 results (78 instances $\times$ 5 tools) independently. We then measured the inter-rater reliability using Cohen's Kappa~\cite{cohensKappa1960}. The Kappa coefficient was $\kappa $ = 0.813, indicating an \textit{almost perfect agreement} between the participants (ref. ~\cite{ cohensKappaInterpretation1977}: $\kappa >$ 0.81). In cases of disagreement, the participants discussed the results with each other to reconcile the differences with one of the authors mediating the process~\cite{AROMA:OOPSLA2019, Monperrus-ESE2017}.

\textbf{Metrics.}
We define two rating scales to evaluate the patch and the post, as shown in \Cref{tab:rating}. Our rating scales are largely inspired from the approach of Zimmermann et al.~\cite{scaleSelection2015, scaleSelection2014}, and follow the advice of Kitchenham et al.~\cite{metricsGuidelines2008} to define a balanced scale and to exclude a ``Don't Know" category if the participants are experts in the field.

The patch and post ratings implicate four possible outcomes of \toolname and its baselines: (1) a correct patch derived from a perfect post, (2) an almost correct patch from a relevant (perfect/helpful) post, (3) no/incorrect patch but a relevant post, and (4) no/incorrect patch and no/irrelevant post. The first three constitute a \textit{useful artifact} in assisting developers for fixing \acp{re}, with (1) being the most desirable output.

\begin{table}[t]\scriptsize
\centering
\setlength\tabcolsep{4pt}
\begin{tabular}{l@{\hskip 2pt}r|cc|cc|c|c}
\toprule
{\bf \ac{re} type} & {\bf \#inst} & \multicolumn{2}{c|}{\bf C$\mathcal{P}$} & \multicolumn{2}{c|}{\bf A$\mathcal{P}$} & {\bf I$\mathcal{P}$} & {\bf I$\mathcal{P}$}\\
& & \multicolumn{2}{c|}{\bf R$\mathbb{P}$} & \multicolumn{2}{c|}{\bf R$\mathbb{P}$} & {\bf R$\mathbb{P}$} & {\bf I$\mathbb{P}$}\\
\midrule
& & {\it Top-1} & {\it Top-3} & {\it Top-1} & {\it Top-3} & & \\
\midrule
ClassCastException & 8 & 5 & 6 & 0 & 0 & 2 & 0 \\
ConcurrentModificationException & 8 & 0 & 0 & 6 & 6 & 2 & 0 \\
IllegalArgumentException & 8 & 1 & 1 & 0 & 0 & 2 & 5 \\
IllegalStateException & 8 & 2 & 2 & 0 & 0 & 0 & 6 \\
IndexOutOfBoundsException & 8 & 1 & 5 & 0 & 1 & 2 & 0 \\
NullPointerException & 8 & 3 & 5 & 0 & 0 & 3 & 0 \\
\midrule
ArithmeticException & 4 & 4 & 4 & 0 & 0 & 0 & 0 \\
NoSuchElementException & 4 & 1 & 2 & 0 & 0 & 0 & 2 \\
RejectedExecutionException & 4 & 0 & 0 & 0 & 0 & 4 & 0 \\
SecurityException & 4 & 0 & 0 & 0 & 0 & 3 & 1 \\
UnsupportedOperationException & 4 & 1 & 1 & 0 & 0 & 2 & 1 \\
\midrule
EmptyStackException & 2 & 1 & 2 & 0 & 0 & 0 & 0 \\
NegativeArraySizeException & 2 & 2 & 2 & 0 & 0 & 0 & 0 \\
\midrule
ArrayStoreException & 1 & 0 & 0 & 0 & 0 & 1 & 0 \\
BufferOverflowException & 1 & 0 & 0 & 0 & 0 & 1 & 0 \\
BufferUnderflowException & 1 & 0 & 0 & 0 & 0 & 1 & 0 \\
CMMException & 1 & 0 & 0 & 0 & 0 & 1 & 0 \\
IllegalMonitorStateException & 1 & 0 & 1 & 0 & 0 & 0 & 0 \\
MissingResourceException & 1 & 0 & 0 & 0 & 0 & 1 & 0 \\
\midrule
{\textbf{Total}} & \textbf{78} & \textbf{21} & \textbf{31} & \textbf{6} & \textbf{7} & \textbf{25} & \textbf{15} \\
\bottomrule
\end{tabular}
\vspace{0.3em}
\begin{tablenotes}
    \centering
    \item[1] {\bf C$\mathcal{P}$}: Correct patch, \hspace{0.2em} {\bf A$\mathcal{P}$}: Almost Correct patch, \hspace{0.2em} {\bf I$\mathcal{P}$}: No/Incorrect patch
    \item[2] {\bf R$\mathbb{P}$}: Perfect/Helpful post, \hspace{0.2em} {\bf I$\mathbb{P}$}: No/Irrelevant post, \hspace{0.2em} {\bf \#inst}: no. of instances
    \vspace{0.25em}
    \item[3] {\footnotesize \textbf{Average runtime per instance = 1 second} (median = 0.6 sec)}
\end{tablenotes}
\caption{Effectiveness Results of \toolname}
\label{tab:rq1}
\end{table}

\begin{figure}[b]
\centering
\begin{subfigure}[b]{\columnwidth}
\begin{lstlisting}[language=Java, framexleftmargin=0.8em, xleftmargin=1.1em, numbersep=3pt, numberstyle=\scriptsize, , firstnumber=1]  
for (String str : new ArrayList<String>(listOfStr)) { @{\hspace{0.5em}\textcolor{javagreen}{\large \cmark}@
    listOfStr.remove(/* object reference or index */); @{\hspace{0.5em}\textcolor{red}{\large \xmark}@
}
\end{lstlisting}
\vspace{-10pt}
\caption{\label{fig:rq1-eg2-so}Stack Overflow post answer \#11201224}
\end{subfigure}
\par\vspace{0.6em}
\begin{subfigure}[b]{\columnwidth}
\begin{lstlisting}[language=Java, framexleftmargin=0.8em, xleftmargin=1.4em, numbersep=3pt, numberstyle=\scriptsize, , firstnumber=1]
@\colorbox{githubRed}{\strut \textbf{-} for (Order order : orders) \{}@
@\colorbox{githubGreen}{\strut \textbf{+} for (Order order : new ArrayList$<$Order$>$orders) \{}@ @{\hspace{0.5em}\textcolor{javagreen}{\large \cmark}@
@\colorbox{githubRed}{\strut \textbf{-} \hspace{1.5em} orders.remove(order);}@
@\colorbox{githubGreen}{\strut \textbf{+} \hspace{1.5em} orders.remove();}@ @{\hspace{0.5em}\textcolor{red}{\large \xmark}@
\end{lstlisting}
\vspace{-10pt}
\caption{\label{fig:rq1-eg2-buggy}\toolname patch for swagger-api/swagger-core}
\end{subfigure}
\caption{\label{fig:rq1-eg2}Almost Correct Patch for ConcurrentModification}
\vspace{-7pt}
\end{figure}

\subsection{RQ1: Effectiveness of \toolname}
\label{sec:rq1}

\Cref{tab:rq1} shows the results of RQ1. Column ``\#inst" shows the number of instances per \ac{re} type. The remaining columns show details of the four possible outcomes of \toolname as discussed in \Cref{sec:protocol}. For example, \textbf{C}$\mathcal{P}$, \textbf{R}$\mathbb{P}$, implies the first outcome: correct patch derived from a relevant post.

\toolname returned a correct repair patch at top-1 position in 27\% of the cases, within the top-3 in 40\% of the cases, and produced an overall useful artifact in 81\% instances. \toolname was fast, requiring an average of 1 second (median = 0.6 sec) end-to-end on a 6-core MacOS laptop. Thus, \toolname can be effective in providing real-time assistance to developers for fixing \acp{re}. 

In a diverse 14 out of 19 \ac{re} types, \toolname returned at least one useful artifact for every instance. \toolname was successful on \acp{re}, such as ClassCastException and IndexOutOfBoundsException, that comprised of commonly occurring exception scenarios and \ac{so} posts that recommended an accurate repair. \acp{re}, such as IllegalArgumentException and IllegalStateException, proved problematic since they are comprised of numerous scenarios that are not a part of our \ac{pattern} library, or scenarios that are very rare or application specific, and hence do not have representation on \ac{so}.

We investigated the results to understand why \toolname was not able to successfully generate a correct repair patch in all useful artifact cases. The first reason is that \toolname produced \textit{almost correct patches} if the fix suggested \ac{so} posts was inadvertently malformed. An example of such a case is shown in \Cref{fig:rq1-eg2} for ConcurrentModificationException which is thrown when a Collection object (\eg\xspace\code{List}) is structurally modified (\eg\xspace\code{remove()}) during iteration. The \ac{so} post suggests a fix of creating a temporary copy of the list and using this new list for iteration (line 1 of \Cref{fig:rq1-eg2-so}). However on line 2, the \ac{so} answer inadvertently omits the argument of \code{remove()}. Therefore, when this fix is translated to the developer's buggy code (\Cref{fig:rq1-eg2-buggy}), it produces a patch that is one token away from the correct patch (\ie missing argument \code{order} in \code{remove()}).

The second reason is when the fix in \ac{so} posts is insufficient in capturing all of the necessary changes to repair the developer's \ac{re}. For example, consider a NullPointerException thrown at line \code{v.m1().m2()}. The correct fix is to guard the failing line with two chained null checks: \code{if(v != null \&\& v.m1() != null) \{...\}}. However, the \ac{so} post only suggests the first null check since its question code snippet discusses the failure of the kind \code{t.foo()}. Such incomplete patches are judged incorrect in our rating, since they are more than one token edit from the complete patch.

\begin{tcolorbox}[left=2pt,right=2pt,top=2pt,bottom=2pt]
\textbf{RQ1:} \textit{\toolname demonstrates strong potential for providing real-time assistance in RE repair -- it generates a correct repair patch at the top in \OurCorrectPatchesTopOne instances, within the top-3 in \OurCorrectPatchesTopThree of the cases, and some useful artifact in \OurUsefulArtifacts of the cases, in only \AvgRuntimeInSec second, on average.}
\end{tcolorbox}
\subsection{RQ2: Comparison with State of the Art}
\label{sec:rq2}

\begin{figure}[t]
  \centering
    \includegraphics[width=\columnwidth]{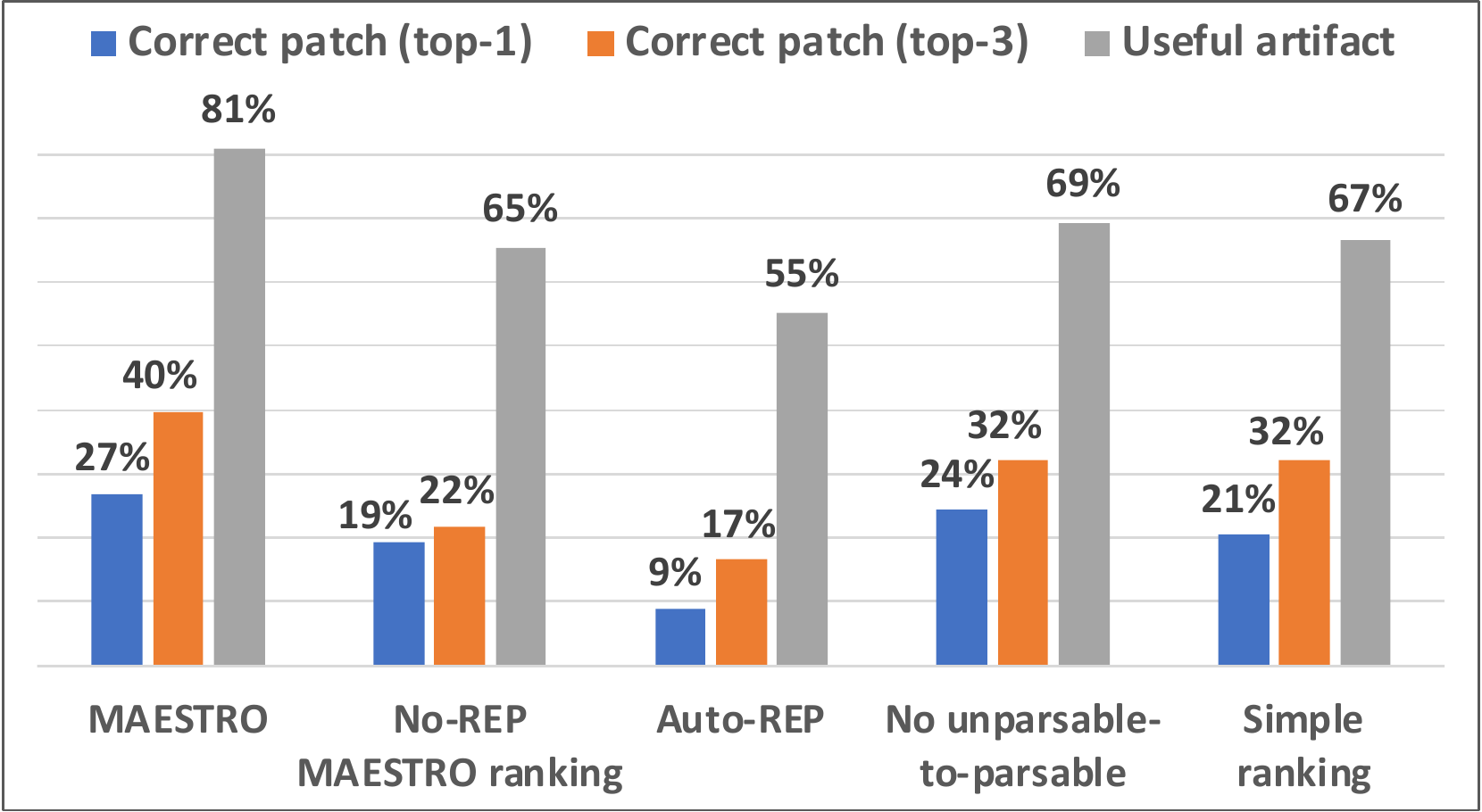}
    \vspace{-15pt}
    \caption{\toolname vs. its baselines}
    \label{fig:rq23}
\end{figure}

A direct comparison with the current state-of-the-art technique, QACrashFix~\cite{QACrashFix:ASE2015}, for generating repair patches using \acl{so} is unfortunately not possible as QACrashFix is designed to generate patches for Android-related crashes using Android-specific information, while our use case is of general-purpose \acp{re}. Instead, we implemented the core algorithm of QACrashFix in our use case, which is to find relevant posts using failure description and transform an \ac{so} fix directly to developer's code without the mediation of a \ac{pattern}.

We made the following changes to \toolname to create this version: (1) removed the mediation of \ac{pattern} from our fixing algorithm, (2) modified the function to find relevant \ac{so} posts to query a web search engine (\eg Google) comprising of the \ac{re} type and failing line from developer's buggy code, since general \acp{re} do not give a Android-like description of the crash. However, in our experiments we found that this search for relevant posts performed rather poorly, ultimately generating a correct repair patch in only 5 instances. Hence, to give this baseline a starting advantage, we provided it with the ranked list of relevant \ac{so} posts from \toolname's indexing. We call this baseline ``\textsc{No-Rep-Maestro-Ranking}".

\Cref{fig:rq23} shows the results for RQ2. \textsc{No-Rep-Maestro-Ranking} returns a correct patch at top-1 position in \NoREPCorrectPatchesTopOne of the cases, within top-3 in \NoREPCorrectPatchesTopThree cases, and an overall useful artifact in \NoREPUsefulArtifacts cases. Thus, it falls significantly short of \toolname on all metrics, producing \NoREPPatchLoss fewer correct patches in the top-3 than \toolname. The primary reason for the performance deficit of \textsc{No-Rep-Maestro-Ranking} is that, even with the benefit of \toolname's post ranking, without the mediation of the \ac{pattern}, it is challenging to precisely map program elements from \ac{so} post's question code snippet to the developer's buggy code. For example, consider the patch generated by \textsc{No-Rep-Maestro-Ranking} shown in \Cref{fig:rq2-eg1} for our illustrative example (\Cref{sec:motivating-example}). \textsc{No-Rep-Maestro-Ranking} simply transfers \code{new String[...]} from the \ac{so} post without any adaptation, leading to an imperfect patch.

\begin{tcolorbox}[left=2pt,right=2pt,top=2pt,bottom=2pt]
\textbf{RQ2:} \textit{A baseline representing state-of-the-art \ac{so}-based patch synthesis produces \NoREPPatchLoss fewer correct patches than \toolname, even when provided with \toolname's ranking.}
\end{tcolorbox}

\begin{figure}[t]
\centering
\begin{lstlisting}[language=Java, framexleftmargin=0.8em, xleftmargin=1.4em, numbersep=3pt, numberstyle=\scriptsize, firstnumber=1]
@\colorbox{githubRed}{\strut \textbf{-} URL[] array = (URL[])urls.toArray();}@
@\colorbox{githubGreen}{\strut \textbf{+} URL[] array = urls.toArray(}@@\colorbox{pastelyellow}{\strut new String}@@\colorbox{githubGreen}{\strut [urls.size()]);}@
\end{lstlisting}
\vspace{-6pt}
\caption{\label{fig:rq2-eg1}Wrong patch by \textsc{No-Rep-Maestro-Ranking} for example in \Cref{fig:example}
(\colorbox{pastelyellow}{\strut Yellow} shows problem in the patch)}
\vspace{4pt}
\end{figure}
\subsection{RQ3: Key Contributions of \toolname}
\label{sec:rq3}

We create three baselines of \toolname. The first baseline, \toolname-\textsc{NoUnparsableToParsable}, measures the impact of our unparsable-to-parsable algorithm (\Cref{sec:so-pool-preparation}) by only using \ac{so} posts with readily available parsable code snippets. The second baseline, \toolname-\textsc{SimpleRanking}, assesses the importance of our \ac{pattern}-based indexing algorithm (\Cref{sec:indexing}) by replacing it with a naive user-votes based ranking of \ac{so} posts and their answers. Finally, our third baseline, \toolname-\textsc{AutoRep}, evaluates the importance of our hand-written \ac{pattern}-library by replacing it with the patterns auto-extracted using Q\&A localization in \toolname's prior version~\cite{MAESTRO:FSE2020}.

\Cref{fig:rq23} shows the results for RQ3. The three baselines return a useful artifact in 55--69\% of the cases (vs. \OurUsefulArtifacts for \toolname), with a correct patch reported in top-3 in only 17--32\% of the cases (vs. \OurCorrectPatchesTopThree for \toolname). Effectively, both the \toolname-\textsc{NoUnparsableToParsable} and  \toolname-\textsc{SimpleRanking} baselines produce 20\% fewer top-3 correct patches than \toolname, showing that unparsable code snippets or 
low-voted posts may at times exclusively contain the correct fixes. The \toolname-\textsc{AutoRep} baseline performs the worst. This is because these auto-\acp{pattern} are post-specific, and may even be sub-optimal if the answer code snippets are lengthy and/or non-specific. Such approximate patterns may be adequate for searching relevant posts (the target of \cite{MAESTRO:FSE2020}), but not for patch generation.

\begin{tcolorbox}[left=2pt,right=2pt,top=2pt,bottom=2pt]
\textbf{RQ3:} \textit{Each of the three components contributes meaningfully to boosting the overall performance of \toolname.}
\end{tcolorbox}

\begin{figure}[t]
  \centering
    \includegraphics[width=0.97\columnwidth]{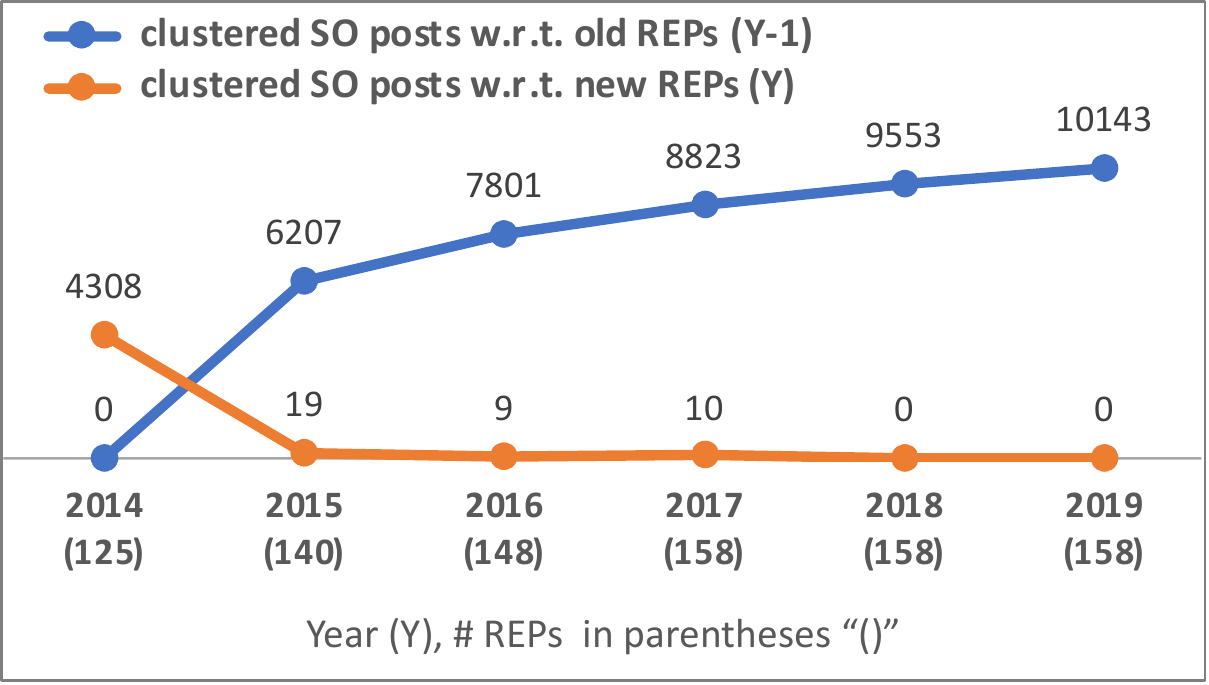}
    \vspace{-4pt}
    \caption{Overhead of maintaining \ac{pattern} library over 5 years}
    \label{fig:rq4}
\end{figure}

\subsection{RQ4: Maintenance Cost for the \ac{pattern} Library}
\label{sec:rq4}

\Cref{fig:rq4} shows the potential cost of compiling and maintaining a \ac{pattern} library over a span of 5 years. For estimating this, we projected back from our \ac{pattern} library curated from the 2019 \ac{so} snapshot by tracking the chronology of the posts, and assuming an average writing time of 5 minutes per \ac{pattern}, calculated based on our 2019 effort. Based on this proposition, compiling the library of 125 \acp{pattern} for the first time in 2014 is a seemingly \textit{inexpensive task}, requiring roughly 10 person-hours. In 2015, almost 2,000 posts could be automatically clustered with the \acp{pattern} from 2014, as they exhibited the same exception-triggering patterns. This demonstrates that writing the \acp{pattern} is indeed a \textit{one-time investment}, with no extra cost thereafter. An \textit{incremental addition} of 15 new \acp{pattern}, clustering 19 \ac{so} posts, came at a significantly low overhead of only 1 person-hour. In subsequent years, the cost-effectiveness of \acp{pattern} is evident from the increased clustering of new posts with old \acp{pattern}, and significantly diminished human effort, with no new \ac{pattern} added after 2017, conceivably because most distinct, popular \acp{pattern} had already been written.


\begin{tcolorbox}[left=2pt,right=2pt,top=2pt,bottom=2pt]
\textbf{RQ4:} \textit{Compiling and maintaining the \ac{pattern} library is a low-cost, incremental, and one-time undertaking.}
\end{tcolorbox}

\subsection{Limitations}
\label{sec:limitations}
\textbf{Repair scope.} \toolname's current implementation can only generate fixes limited to intra-procedural exception scenarios. This design choice is motivated by the observation that \ac{re} scenarios generally tend to be concise and local in nature. However, to expand the scope, we plan to extend \toolname to handle inter-procedural analysis in the future.

\textbf{Usefulness of artifacts.} Judging repair patches and \ac{so} posts is a subjective task that is performed manually, posing a threat to construct validity. Manual judgement for correctness of patches is an establishednorm in the \ac{apr} community (\eg \cite{elixir2017, cocount:ISSTA2020, APREvaluation:JSS2021}). Nonetheless, to minimize this threat, we clearly defined the criteria for our rating scales, consistent with prior work~\cite{scaleSelection2015, scaleSelection2014, metricsGuidelines2008, MAESTRO:FSE2020}. Further, to further reduce bias, we recruited two \textit{external participants} to evaluate independently. We then used Cohen's Kappa~\cite{cohensKappa1960} to measure inter-rater reliability, which showed \textit{almost perfect agreement} among the participants. Disagreements were reconciled via discussion -- consistent with previous work~\cite{Monperrus-ESE2017, AROMA:OOPSLA2019, MAESTRO:FSE2020}.

\textbf{Mining \acp{pattern}} \toolname employs a semi-automated process for extracting \acp{pattern} from \ac{so} posts. As we show through experiments in RQ3, automatically extracting \acp{pattern} is challenged by the sub-optimal Q\&A localization. To make the mining process efficient, \toolname only suggests a small number of \ac{so} posts to the human that are likely to represent diverse \acp{pattern}, and also aids the writing task by presenting an auto-extracted approximate \ac{pattern}. Further, compiling a library of \acp{pattern} through our mining process is a one-time undertaking that could be performed incrementally. 
\section{Related Work}
\label{sec:relatedWork}

\textbf{\ac{apr} using \ac{so}.} 
QACrashFix~\cite{QACrashFix:ASE2015} is closely related to our work. QACrashFix is a generate-and-validate \ac{apr} technique for fixing Android-related crash bugs. As shown in \Cref{sec:evaluation}, QACrashFix does not work well for our use case of fixing general-purpose \acp{re} for two reasons. First, QACrashFix is limited in finding relevant \ac{so} posts for general \acp{re}, since they often do not have a crash description that QACrashFix uses to do the search. Second, even when provided with \toolname's \ac{so} posts ranking, it fails to accurately generate a correct patch in 45\% of the cases. Different from our use case of generating repairs from \ac{so} posts, another \ac{apr} technique, SOFix~\cite{SOFix:SANER2018}, manually extracts a set of repair templates from \ac{so}.

\textbf{Assisting debugging of \acp{re}.} 
FuzzyCatch~\cite{FuzzyCatch:FSE2020} recommends Android-related exception handling (\code{try-catch} blocks). Several techniques provide tailored repair solutions for only the common \acp{re}, such as NullPointerException (NPE). Sinha et al.~\cite{Sinha:ISSTA2009} use stack traces to locate and create a patch, NPEFix~\cite{NPEFix:arXiv2015} uses NPE-specific heuristics, VFix~\cite{VFix:ICSE2019} uses data and control-flow analysis to prune the NPE repair space, Getafix~\cite{GetAFix:OOPSLA:2019} and Genesis~\cite{Genesis:FSE2017} learn fix patterns from  human-written patches, and Droix~\cite{Droix:ICSE2018} uses search-based algorithms. Another research isolates and recovers from runtime errors~\cite{Long:PLDI2014, Ares:ASE2016}, yet another focuses on the automatic generation of test oracles for \acp{re}~\cite{Toradocu:ISSTA2016}. By contrast, our work presents a general-purpose repair approach that applies to a diverse set of \acp{re} by leveraging crowd intelligence encoded in forums (\eg \ac{so}).

\textbf{\acf{apr}.}
We share the patch-generation goal of traditional \ac{apr} techniques. Generate-and-validate \ac{apr} techniques, such as SketchFix~\cite{SketchFix:FSE2018}, Kali~\cite{Kali:ISSTA2015}, SPR~\cite{SPR:FSE2015}, Elixir~\cite{elixir2017}, Hercules~\cite{hercules2019}, Angelix~\cite{Angelix:ICSE2016}, and GenProg~\cite{GenProg:ICSE2012}, explore a search space of manually crafted repair transformations that are tried in succession until a plausible repair is found. Another group of techniques, such as CoCoNut~\cite{cocount:ISSTA2020}, Sequencer~\cite{SEQUENCER:TSE2019}, and DLFix~\cite{DLFix:ICSE2020}, use deep learning to auto-learn the repair transformations. Yet other techniques, such as Phoenix~\cite{Phoenix:FSE2019} and Refazer~\cite{Refazer:ICSE2017}, use programming by example to learn repair strategies. Unlike us, such traditional \ac{apr} techniques typically rely on a dynamic patching oracle (\eg test suite) and thereby involve long running times, making them unfit for our use case of real-time assistance when patching oracles may not even be available.

\textbf{Mining developer forums (\eg \ac{so}).}
The previous version of \maestro~\cite{MAESTRO:FSE2020, maestro-tool:ASE2021} recommends relevant \ac{so} posts for manual fixing of \acp{re}, while Prompter~\cite{Prompter:MSR2014, Seahawk:CSMR2013} and Libra~\cite{Libra:ICSE2017} suggest posts to assist during implementation. ExampleStack~\cite{ExampleStack:ICSE2019} shows examples of \ac{so} adaptations. FaCoY~\cite{FaCoY:ICSE2018} performs code-to-code search. Nagy et al.~\cite{Nagy:ICSME2015} mine common SQL error patterns. AnswerBot~\cite{AnswerBot:ASE2017} and Crokage~\cite{Crokage:ICPC2019} summarize \ac{so} answers. Chen et al.~\cite{CrowdDebuggingFSE2015} use \ac{so} to fault localize code and suggest posts. SEQUER~\cite{SEQUER:ICSE2021} reformulates user queries to find posts. CSnippEx~\cite{CSNIPPEX:ISSTA2016} makes non-compilable \ac{so} code snippets compilable. By contrast, such techniques are not related to our use case of \ac{so}-based patching.
\section{Conclusion}
We extend our prior work, \toolname, by adding real-time patching support for fixing \acp{re} using \ac{so} posts. \toolname exploits a library of \acp{pattern} semi-automatically mined from \ac{so} posts through a one-time, incremental process. A \ac{pattern} represents an \ac{re} pattern and is used to mediate each of the key steps: indexing \ac{so} posts, retrieving a relevant post matching the \ac{re} scenario exhibited by the developer's code, and finally adapting the post-suggested answer to fix developer's buggy code. An evaluation on a published benchmark of 78 \ac{re} instances showed that \toolname generated a correct repair patch at top-1 in \OurCorrectPatchesTopOne of the cases, within the top-3 in \OurCorrectPatchesTopThree cases, and an overall useful artifact in \OurUsefulArtifacts cases. Further, \toolname only required \AvgRuntimeInSec second, on average, per instance.

\newpage
\balance
\bibliographystyle{IEEEtran}
\bibliography{IEEEabrv,runtime}

\end{document}